\documentclass[journal]{IEEEtran}
\PassOptionsToPackage{svgnames,table}{xcolor}
\usepackage{amsmath,amssymb,graphicx,multirow,array,enumitem,adjustbox}
\usepackage{booktabs, threeparttable,caption, makecell}
\usepackage{stfloats}
\usepackage{makecell}
\usepackage{pifont,wasysym}
\usepackage{utfsym}
\usepackage{bbm}
\usepackage[ruled, lined, linesnumbered, commentsnumbered, longend]{algorithm2e}
\usepackage[colorlinks=true,citecolor=green]{hyperref}
\usepackage{cleveref}
\usepackage{color}
\usepackage{subcaption}
\usepackage{adjustbox}

\ifCLASSINFOpdf
\else
\fi

\begin{document}

\title{Exploring Self-Supervised Audio Models for Generalized Anomalous Sound Detection}

\author{Bing Han, \IEEEmembership{Student Member, IEEE,}
        Anbai Jiang, 
        Xinhu Zheng, 
        Wei-Qiang Zhang, 
        Jia Liu, \\
        Pingyi Fan, 
        and Yanmin Qian, \IEEEmembership{Senior Member, IEEE}
\thanks{
This paper is an extension based on ICASSP 2024~\cite{han2024exploring} and Interspeech 2024~\cite{jiang24c_interspeech}.

Bing Han, Xinhu Zheng and Yanmin Qian are with the AudioCC Lab, Department of Computer Science and Engineering \& MoE Key Laboratory of Artificial Intelligence, AI Institute, Shanghai Jiao Tong University, Shanghai, 200240 P. R. China (e-mail:\{hanbing97, zhengxh24, yanminqian\}@sjtu.edu.cn).

Anbai Jiang, Wei-Qiang Zhang, Pingyi Fan, and Jia Liu are with Department of Electronic Engineering, Tsinghua University, Beijing, 100084 P. R. China (e-mail:jab22@mails.tsinghua.edu.cn,\{wqzhang, fpy, liuj\}@tsinghua.edu.cn).

}
}

\markboth{Journal of \LaTeX\ Class Files,~Vol.~14, No.~8, August~2015}%
{Shell \MakeLowercase{\textit{et al.}}: Bare Demo of IEEEtran.cls for IEEE Journals}
%

\maketitle

\begin{abstract}
Machine anomalous sound detection (ASD) is a valuable technique across various applications. However, its generalization performance is often limited due to challenges in data collection and the complexity of acoustic environments. Inspired by the success of large pre-trained models in numerous fields, this paper introduces a robust ASD model that leverages self-supervised pre-trained models trained on large-scale speech and audio datasets. Although there are inconsistencies between the pre-training datasets and the ASD task, our findings indicate that pre-training still provides substantial benefits for ASD.
To mitigate overfitting and retain learned knowledge when fine-tuning with limited data, we explore Fully-Connected Low-Rank Adaptation (LoRA) as an alternative to full fine-tuning. Additionally, we propose a Machine-aware Group Adapter module, which enables the model to capture differences between various machines within a unified framework, thereby enhancing the generalization performance of ASD systems.
To address the challenge of missing attribute labels, we design a novel objective function that dynamically clusters unattributed data using vector quantization and optimizes through a dual-level contrastive learning loss.
The proposed methods are evaluated on all benchmark datasets, including the DCASE 2020-2024 five ASD challenges, and the experimental results show significant improvements of our new approach and demonstrate the effectiveness of our proposed strategies.

\end{abstract}


\IEEEpeerreviewmaketitle

\section{Introduction}

\IEEEPARstart{A}{nomalous} sound detection (ASD)~\cite{koizumi2020description} focuses on determining whether sounds generated by a specific machine are normal or anomalous. This research area has attracted significant attention because of its critical role in improving the safety and operational efficiency of industrial machinery. 
However, anomalous sound detection models often face the problem of generalization in practical applications. 
Firstly, due to the challenges associated with collecting anomalous samples, ASD usually operates within an unsupervised learning framework, relying solely on normal-state samples for training. The objective of ASD is to assess whether a given query sample belongs to the anomaly class, which encompasses a variety of potential abnormal conditions. 
Secondly, limited by application scenarios, it is often required that the model can quickly generalize to unseen machine types and be robust to acoustic environments. 
Considering these challenges, DCASE has held challenge competitions for many years~\cite{koizumi2020description, kawaguchi2021description, dohi2022description, dohi2023description, nishida2024description}, and many participants have also proposed many valuable technical solutions for generalized ASD.

Traditionally, machine ASD has relied on mechanism-based approaches that utilize the physical characteristics or operational principles of machines to detect abnormal sounds. Although effective in certain scenarios, these methods are constrained by their dependence on domain-specific knowledge and their limited extension to diverse acoustic environments. Recently, however, there has been a marked shift towards deep learning approaches for ASD, which can be broadly categorized into reconstruction-based and self-supervised-based methods~\cite{ruff2021unifying}. Reconstruction-based methods~\cite{suefusa2020anomalous, dohi2021flow} aim to improve ASD efficiency by modeling the distribution of normal sounds, then identifying anomalies by evaluating how likely unknown sounds fit within this distribution. In contrast, self-supervised methods~\cite{Giri2020, jiang2023unsupervised, chen2022self} often utilize metadata from audio files, such as machine types and operational statuses, as pseudo-labels to train classifiers that capture latent representations of machine sounds.
Despite advances in these methods, their generalizability remains limited. This limitation mainly arises from the complex nature of diverse acoustic environments, the difficulties involved in data collection, and the scarcity of attribute-specific information.

In the fields of speech and audio processing, leveraging large-scale pre-trained models has become the dominant approach for achieving state-of-the-art performance, particularly to address the challenge of limited labeled data. Building on the pioneering success of models like BERT~\cite{devlin2018bert} in natural language processing, researchers have developed a range of innovative architectures for masked audio modeling, which apply self-supervised learning to large-scale unlabeled data. These methods have demonstrated notable effectiveness across various speech and audio tasks~\cite{yang2021superb,beats}, showcasing the potential of pre-training to harness vast amounts of unlabeled data in these domains.

In the context of the anomaly sound detection (ASD) task, our objective is to identify robust audio encoders capable of generalizing well, thereby reducing the risk of overfitting due to limited training data. Inspired by the successful application of pre-trained models in diverse downstream tasks~\cite{chen2022large}, it is valuable to explore whether large-scale pre-trained models from related domains can be adapted to ASD tasks, potentially mitigating data scarcity issues despite the domain mismatch between general audio data and machine sounds.

In our previous works~\cite{han2024exploring,jiang24c_interspeech}, we preliminarily explored fully fine-tuning pre-trained speech and audio models for anomalous sound detection (ASD), achieving promising results—ranking 2$^{nd}$ in DCASE 2023 and 1$^{st}$ in DCASE 2024.
In this paper, to further validate the effectiveness of fine-tuning pre-trained models for ASD tasks, we aim to comprehensively explore and compare a broader range of large-scale pre-trained models on ASD tasks, conduct in-depth analyses, and evaluate our system across multiple benchmarks.
However, we observe that fine-tuning self-supervised pre-trained models tends to overfit, limiting their generalization to unseen machine types.
To mitigate these issues and enhance performance, we propose several strategies, including a fully connected multi-branch LoRA and a machine-aware group adapter, to improve generalizability across diverse acoustic environments and unseen machine types.
Finally, in response to the challenge of missing labels in real-world scenarios raised in DCASE 2024~\cite{nishida2024description}, we design a new objective function to compensate for the lack of attribute information.
By integrating these techniques, we construct a novel ASD framework based on self-supervised pre-trained models, which achieves state-of-the-art performance on various benchmarks.

Our contributions can be summarized as follows:
\begin{itemize}
    \item In order to build a robust and generalized ASD model, following our previous works~\cite{han2024exploring,jiang24c_interspeech}, we comprehensively explore several large-scale pre-trained speech or audio models for the ASD task at the first time, and find that the pre-training features can bring effective gains especially on unseen condition, even though the pre-training data include nearly no machine sound.
    \item To avoid knowledge forgetting during the fine-tuning process, we propose Fully-Connected Multi-branch LoRA in the ASD architecture, which can retain effective pre-training representations while also having stronger modeling capabilities of anomalous characteristics.
    \item Meanwhile, to generalize the ASD model to unseen machine types, we propose the Machine-aware Group Adapter module, which can group the differences among different machines in a single model, thereby improving the generalization performance of ASD systems.
    \item To alleviate the negative impact of missing attribute labels, based on contrastive learning, we design a new loss function that can dynamically cluster pseudo labels using a quantizer for optimization.
    \item With these strategies applied, we can build a robust and generalized ASD system, and achieve the \textbf{state-of-the-art} (SOTA) performance across all the ASD benchmarks, including five ASD challenges of DCASE 2020-2024~\cite{koizumi2020description, kawaguchi2021description, dohi2022description, dohi2023description, nishida2024description}.
\end{itemize}

\section{Background}
\subsection{Machine Anomalous Sound Detection}
Inspired by advances in deep learning, recent research has shifted towards data-driven methods for the machine anomaly sound detection task. Deep learning models, particularly convolutional neural networks (CNNs) and transformer~\cite{transformer}, have shown impressive capability in learning complex audio features directly from raw or transformed sound data, bypassing the need for manual feature engineering.

Machine learning-based approaches for ASD can generally be categorized into reconstruction- and discriminative-based methods. Reconstruction methods typically assume that anomalous samples exhibit a larger reconstruction error when trained on normal samples~\cite{suefusa2020anomalous}, or they model the normal distribution directly to estimate the likelihood of data~\cite{dohi2021flow}. In contrast, discriminative methods leverage meta-information for classification optimization, thereby enhancing the model's ability of learning robust audio representations~\cite{Giri2020, jiang2023unsupervised, chen2022self}. 
Among these methods, discriminative approaches are currently more widely adopted in the research community, as they often yield superior performance. Notably, the winning systems in the past three DCASE challenges were all based on discriminative methods~\cite{dohi2022description,dohi2023description,nishida2024description}, and the majority of recent publications have also followed this paradigm~\cite{wilkinghoff2021sub,yin2025diffusion,jiang2025adaptive,guan2025disentangling}.
Some researchers have even combined both approaches in an effort to extract more comprehensive and robust features for anomaly detection~\cite{zeng2023joint}. Furthermore, a variety of data augmentation techniques have been proposed to increase the diversity of training data and mitigate the challenges of data scarcity~\cite{chen2023sw, zhang23fa_interspeech}. 
After developing an anomaly sound encoder, back-end detectors are still required to obtain anomaly scores. Commonly used algorithms for detecting outliers include K-Nearest Neighbors (KNN)~\cite{ramaswamy2000efficient}, Local Outlier Factor (LOF)~\cite{breunig2000lof}, and Gaussian Mixture Models (GMM)~\cite{zong2018deep}.
At present, these methods mainly focus on mining internal features of the dataset, and there have been no attempts to introduce external knowledge to assist, such as adopting models pre-trained on large-scale audio data.

\begin{figure*}[t]
    \centering
    \includegraphics[width=1.0\linewidth]{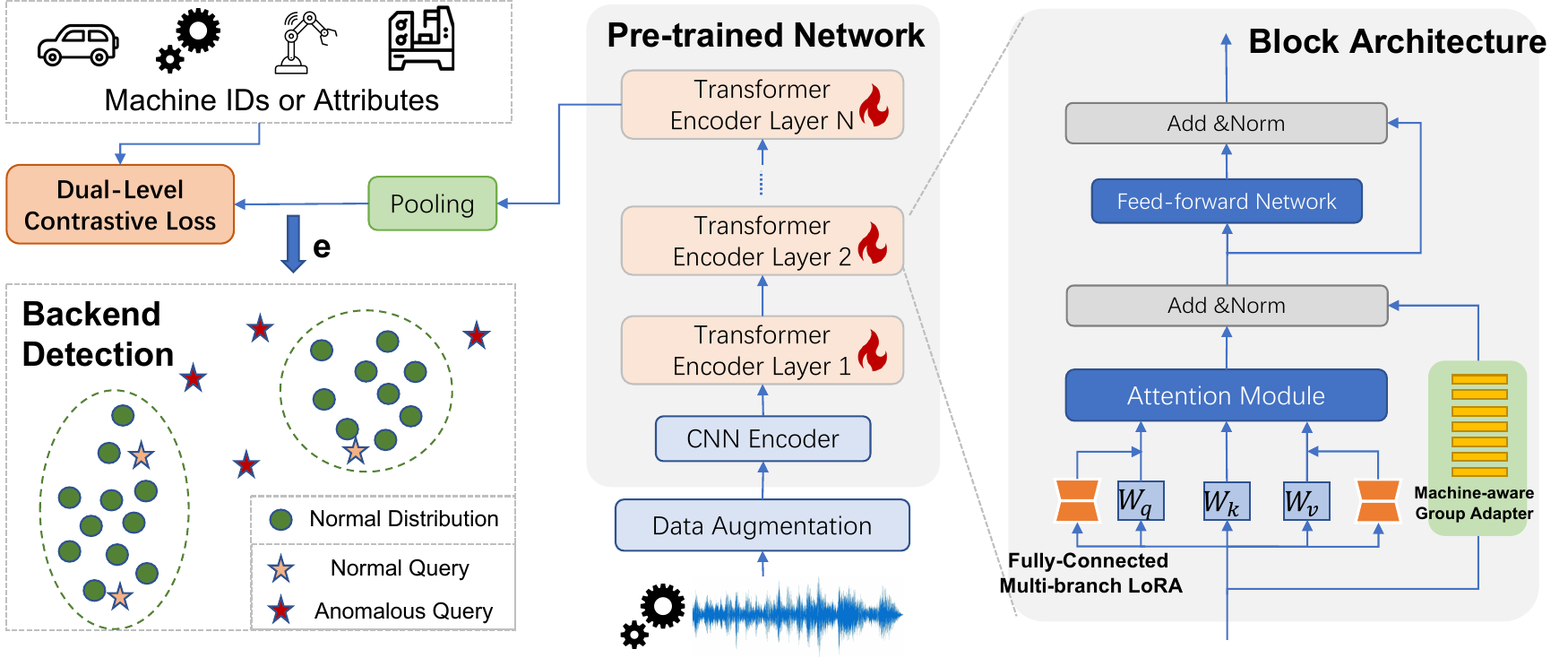}
    \caption{
    Overview of our proposed framework, which fine-tunes audio or speech pre-trained models using the proposed Fully-Connected Multi-Branch LoRA (Sec.~\ref{sec:method_lora}), Machine-Aware Group Adapter (Sec.~\ref{sec:method_adapter}), and Dual-Level Contrastive Loss (Sec.~\ref{sec:method_dlcl}) for the anomalous sound detection task. The fire icon indicates modules that are optimized through backpropagation. The Attention Module and Feed-Forward Network refer to the multi-head attention mechanism and the position-wise feed-forward network, respectively, as introduced in the Transformer architecture~\cite{transformer}. During inference, if the query $\mathbf{e}$ falls within the normal distribution (represented by the green dotted circle), it is classified as normal; otherwise, it is identified as anomalous.
    }
    \label{fig:finetune}
\end{figure*} 

\subsection{Large-Scale Pre-trained Audio and Speech Models}
Large-scale pre-trained audio and speech models have significantly transformed various speech and audio processing tasks, benefiting from vast amounts of data and advanced model architectures. By utilizing self-supervised learning, these models have demonstrated that it is possible to learn rich representations of audio features without massive labeled datasets. These models, typically pre-trained on large-scale audio corpora, have shown remarkable versatility and adaptability, providing strong baselines and fine-tuning capabilities for downstream tasks such as automatic speech recognition (ASR)~\cite{baevski2020wav2vec}, speaker verification~\cite{chen2022large}, audio classification~\cite{beats}, and so on.

For models focusing on speech-related tasks, Wav2Vec 2.0~\cite{baevski2020wav2vec} quantizes raw waveforms into discrete units, which are then modeled using a transformer-based encoder. HuBERT~\cite{hsu2021hubert} generates pseudo-labels by clustering mel spectrograms and adopts the architecture of Wav2Vec 2.0. UniSpeech~\cite{wang2021unispeech} combines contrastive learning with supervised learning, while WavLM~\cite{chen2022wavlm} improves upon HuBERT by incorporating multiple data augmentation strategies.

For models aimed at audio-related tasks, AST~\cite{gong21b_interspeech} fine-tunes a Vision Transformer (ViT)\cite{vit} model, originally pre-trained on image data, for audio classification. BEATs\cite{beats} trains a ViT backbone along with an acoustic tokenizer that generates pseudo-labels for unlabeled data. ImageBind~\cite{girdhar2023imagebind} employs a ViT encoder to align audio with multiple modalities. Recently, ATST~\cite{li2024self} introduced self-teaching techniques for the unsupervised training of ViT models. Compared to the speech models that typically process speech on the frame-level, in contrast, the audio models generally operate on audio in the form of patches.

\subsection{Fine-tuning Methods}
Fine-tuning self-supervised models for downstream tasks typically involves adjusting model parameters with a small amount of labeled data and a task-specific loss function. Usually, self-supervised models consist of a CNN encoder followed by transformer-based encoders~\cite{baevski2020wav2vec,hsu2021hubert,chen2022wavlm}. Several fine-tuning methods are employed to balance performance and parameter efficiency. Full fine-tuning updates all model parameters, often leading to high performance but at the cost of parameter efficiency~\cite{pepino2021emotion}. 
Weight tuning, on the other hand, freezes most of the model while learning a weighted sum of the encoder features, offering better efficiency but often reducing performance~\cite{han2024exploring,chen2022large}. 
LoRA tuning introduces low-rank matrices into the self-attention layers to enable more efficient adaptation with minimal performance loss~\cite{liu2024sparsely}. 
Prefix tuning appends learnable pseudo tokens to encoder layers, facilitating task-specific adjustments, and is more commonly used in generation tasks~\cite{li2021prefix}. Lastly, efficient adapter tuning adds modular adapter components to the model, providing parameter-efficient fine-tuning while retaining the model stability~\cite{thomas2022efficient}. 
These methods offer different trade-offs between parameter efficiency and performance, depending on the task and architecture.

\section{Self-supervised Audio Modeling for Generalized Anomalous Sound Detection}

In this section, we introduce our proposed self-supervised audio pre-trained model-based framework for generalized anomalous sound detection.
First, the overall framework is detailed in Sec.\ref{sec:method_pretrain}.
To address the issue of knowledge forgetting during fine-tuning, we present a novel Fully-Connected Multi-Branch Low-Rank Adaptation method in Sec.\ref{sec:method_lora}.
To further enhance the model’s generalization across different machine types, the Machine-Aware Group Adapter is introduced in Sec.\ref{sec:method_adapter}.
Finally, to mitigate the impact of missing labels in real-world scenarios, a newly designed dual-level contrastive loss function is described in Sec.\ref{sec:method_dlcl}.
Together, these components constitute a robust and generalized framework for anomalous sound detection.

\subsection{Framework overview of Pre-trained Models for Anomalous Sound Detection}
\label{sec:method_pretrain}
Fig.~\ref{fig:finetune} illustrates the overall architecture of our framework, encompassing both the training and detection processes.
In the proposed ASD framework, a Transformer-based pre-trained model equipped with a pooling layer is employed to extract segment-level representations $\mathbf{e}$ from machine audio samples.
Given the distinct characteristics of speech and general audio, the modeling units used in these pre-trained models are accordingly designed to differ.

\textbf{Speech pre-trained models}, the input to the model is typically the raw waveform. Various augmentation strategies are first applied, followed by a 1D convolutional encoder that segments the sequential audio data into continuous \textbf{frame-level} representations along the time axis. Each frame corresponds to a short snippet of the original waveform within a very brief time window and serves as the basic processing unit. This design choice aligns with the training paradigm of many speech pre-trained models, which are often trained on datasets such as LibriSpeech~\cite{panayotov2015librispeech}, with a focus on modeling semantic information.
 
\textbf{Audio pre-trained models}, the waveform is typically converted into a mel-spectrogram as the model input, which is then processed by a 2D convolutional encoder into non-overlapping patch-level latent representations across both spatial and temporal dimensions. Each patch corresponds to a local region of the original spectrogram and is not constrained by strict temporal continuity.
This approach is well-suited for audio pre-trained models, which are often trained on large-scale datasets such as AudioSet~\cite{gemmeke2017audio} to support complex audio analysis tasks.

After applying different processing strategies for speech and audio, the resulting features are flattened into sequences and fed into Transformer blocks, initialized with weights from the corresponding pre-trained models.
Notably, we integrate the proposed Fully-Connected Multi-Branch LoRA and Machine-Aware Group Adapter to enhance the capacity of these Transformer blocks.
Following $N$ layers of Transformer encoding, we apply attentive statistical pooling~\cite{okabe2018attentive} to aggregate the sequence into a fixed-length, utterance-level embedding.
Finally, this embedding is projected through a linear layer to a lower-dimensional representation $\mathbf{e}$, which is subsequently used for anomaly detection.

For auxiliary training objectives, our framework also adopts a discriminative approach to differentiate between machine types or operating conditions as proxy tasks for model optimization, following previous studies~\cite{Giri2020, jiang2023unsupervised, chen2022self}. This choice is motivated by the current dominance of discriminative methods in the field, as evidenced by the fact that the winning systems in the past three DCASE challenges were all based on discriminative approaches~\cite{dohi2022description,dohi2023description,nishida2024description}.
When labels are available, we apply a standard classification loss. In cases where labels are missing, we employ our proposed dual-level contrastive loss to guide model optimization.

In the anomaly detection process, normal sound data will be modeled as normal distributions(green dotted circle). For a query representation $\mathbf{e}$, if it conforms to a normal distribution (pink star in Fig.~\ref{fig:finetune}), it is judged as normal. If it is an outlier (red star in Fig.~\ref{fig:finetune}), it will be judged as anomalous.

\subsection{Fully Connected Multi-Branch Low-rank Adaptation}
\label{sec:method_lora}
\begin{figure}[h]
    \centering
    \includegraphics[width=0.9\linewidth]{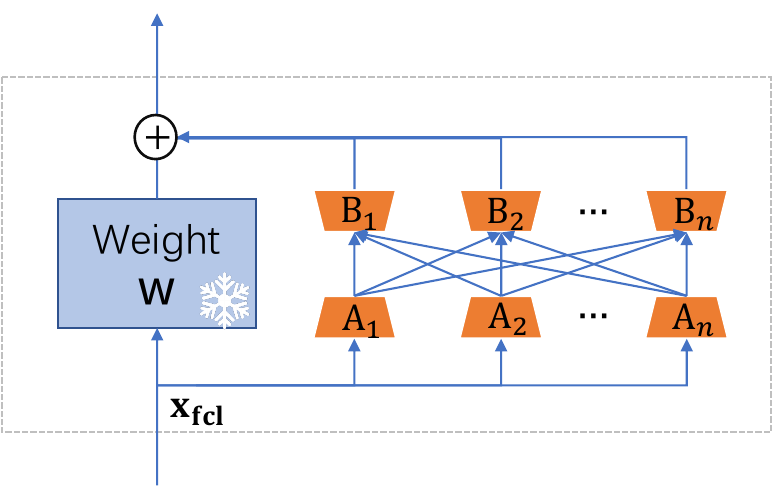}
    \caption{Overview of our proposed fully-connected multi-branch low-rank adaptation structure when fine-tuning pre-trained models for ASD. The snowflake icon denotes that the weight is frozen without updating.}
    \label{fig:fc_lora}
\end{figure}
Low-Rank Adaptation (LoRA)~\cite{hu2021lora} is a widely used technique for adapting the weights of large models to new datasets or tasks without modifying the original architecture. This method is particularly advantageous, as it greatly reduces the computational resources needed for fine-tuning by introducing a small number of free parameters to each Transformer layer, while keeping all original model parameters frozen.
Specifically, for each weight matrix $\mathbf{A} \in \mathbb{R}^{d\times k}$ in a Transformer layer, two new matrices $\mathbf{A} \in \mathbb{R}^{d\times r}$ and $\mathbf{B} \in \mathbb{R}^{r\times k}$ are added, where $d$ is input feature dimension, $k$ is output feature dimension and $ r \ll \min\{d,k\}$ is low-rank bottleneck dimension. 
During training, each matrix multiplication involves the input $\mathbf{x_{fcl}}$ being multiplied with both the original weight matrix $\mathbf{W}$ and the low-rank approximation matrices $\mathbf{A}$ and $\mathbf{B}$. The outputs are then summed to form the final result for further computation. Only the matrices $\mathbf{A}$ and $\mathbf{B}$ are updated during fine-tuning, while $\mathbf{W}$ remains fixed, significantly reducing memory usage.
Furthermore, once fine-tuning is complete, this additional branch can be merged into the original weights, ensuring no extra parameters are introduced during inference:
\begin{align}
\label{equ:lora}
\mathbf{W}\mathbf{x_{fcl}} + \mathbf{BA} \mathbf{x_{fcl}} & = (\mathbf{W}+\mathbf{BA})\mathbf{x_{fcl}}  \\
&= (\mathbf{W} + \Delta \mathbf{W}) \mathbf{x_{fcl}}
\end{align}
where $\Delta \mathbf{W} = \mathbf{BA}$ can be added to the original weights $\mathbf{W}$.

Due to the substantial differences between speech\&audio-based pre-trained models and anomaly sound detection (ASD) tasks, using a single LoRA during fine-tuning can help prevent knowledge forgetting in large models. Moreover, considering the single LoRA approach may also lead to limited adaptation capacity,
we extend the single LoRA structure ${\mathbf{A}, \mathbf{B}}$ into a parallel multi-branch LoRA framework, represented as $\{\mathbf{A}_1, \mathbf{A}_2, \dots, \mathbf{A}_{n_l}, \mathbf{B}_1, \mathbf{B}_2, \dots, \mathbf{B}_{n_l}\}$, where $n_l$ denotes the number of branches. To further facilitate information exchange among these branches, we incorporate interconnected structures inspired by fully connected layers (FCL). The overall architecture of the proposed fully connected multi-branch LoRA is illustrated in Fig.~\ref{fig:fc_lora}. Specifically, for the $i$-th branch, the input to $\mathbf{B}_i$ is derived from the aggregation of $\mathbf{A}_1$ through $\mathbf{A}_{n_l}$, while the outputs of $\mathbf{B}_1$ to $\mathbf{B}_{n_l}$ are also aggregated and merged into the main branch:
\begin{align}
    \mathbf{W}\mathbf{x_{fcl}} + \sum_{i=1}^{n_l}\mathbf{B}_i(\sum_{j=1}^{n_l}\mathbf{A}_j \mathbf{x_{fcl}}) &= (\mathbf{W} + \sum_{i=1}^{n_l}\sum_{j=1}^{n_l}\mathbf{B}_i\mathbf{A}_j) \mathbf{x_{fcl}} \\
    & = (\mathbf{W} + \Delta \mathbf{W}) \mathbf{x_{fcl}}
\end{align}
where all the branches can also be merged with $\Delta \mathbf{W} = \sum_{i=1}^{n_l}\sum_{j=1}^{n_l}\mathbf{B}_i\mathbf{A}_j$, without introducing any additional parameters for inference, while the multi-branches LoRA can improve the adaptability to the ASD task during training.

\subsection{Machine-Aware Group Adapter}
\label{sec:method_adapter}
\begin{figure}[ht]
    \centering
    \renewcommand{\thefigure}{3}
    \includegraphics[width=0.9\linewidth]{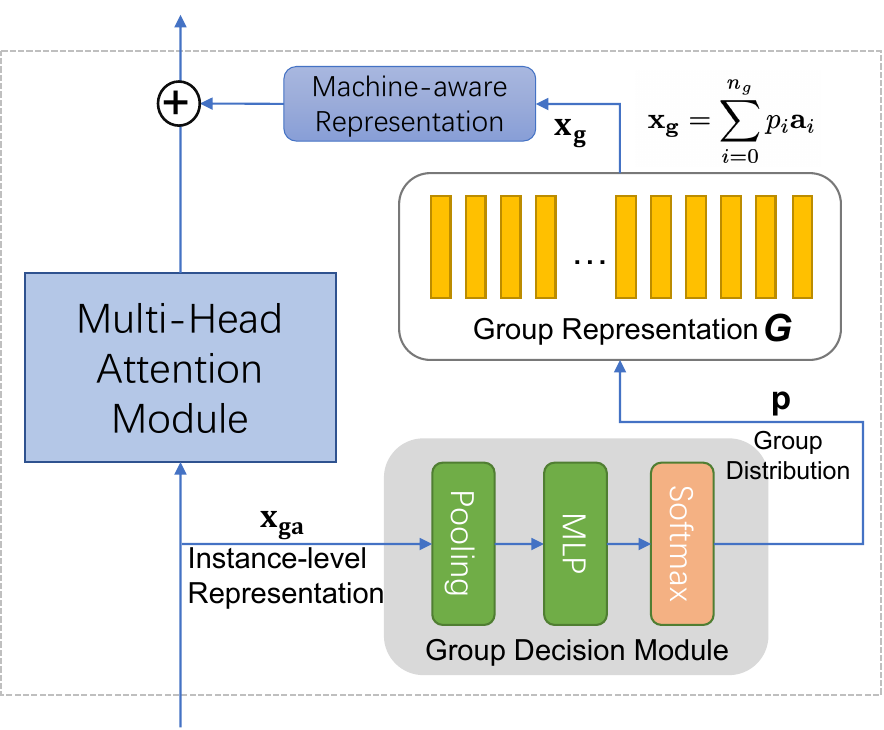}
    \caption{Overview of our proposed group adapter, which is inserted into the main branch for modeling the differences among the multiple machines. For the forward process, the instance-level representation $\mathbf{x_{ga}}$. are fed into the group decision module and obtain the group distribution $\mathbf{p}$. Then, it will be multiplied by the group representation set $\boldsymbol{G}$, weighted summed to obtain a machine-aware representation $\mathbf{x_g}$, and finally added to the main branch.
    It's noted that the vectors in the group representation are normalized during the optimization to keep their modulus equal to 1.}
    \label{fig:groupadapter}
\end{figure}
Discriminative-based methods, which represent the current mainstream approaches for anomalous sound detection (ASD), typically train audio encoders using softmax-based loss functions to extract instance-level representations. However, such representations may lack the capacity to capture machine-specific characteristics~\cite{LiuCQUPT2022}.
To address this limitation, we propose a Machine-aware Group Adapter module, illustrated in Fig.~\ref{fig:groupadapter}, which is added to enhance multi-head attention module in transformer layers. 
Specifically, the latent representation $\mathbf{x_{ga}}$ from the backbone network serves as the instance-level feature, which is then fed into the group decision module $g(\cdot)$. This module comprises an average pooling layer across the time dimension, followed by a MLP (multi-layer, composed of multiple linear layers) and a softmax layer to predict the probability across $n_g$ groups, yielding $\mathbf{p}=\{p_0, p_1, \dots, p_i, \dots, p_{n_g}\}$, where $n_g$ denotes the number of group adapters and $p_i$ denotes the probability of $i$-th group representation $\mathbf{a}_i$. The distribution $\mathbf{p}$ can be calculated as follows:
\begin{equation}
    \label{equ:softmax}
    \mathbf{p} = \mathrm{Softmax}(g(\mathbf{x_{ga}})/\tau_1)
\end{equation}
$\tau_1$ is the temperature factor, which can sharpen the shape of distribution and avoid collapsing into the same group.

Subsequently, based on the obtained group probability $\mathbf{p}$, we can weighted sum up the group representation set $\boldsymbol{G} = \{\mathbf{a}_1, \mathbf{a}_2, \dots, \mathbf{a}_i,\dots, \mathbf{a}_{n_g}\}$ with Equ.~\ref{equ:weight_sum} across $n_q$ groups:
\begin{equation}
\label{equ:weight_sum}
\mathbf{x_{g}} =  \sum_{i=0}^{n_g} p_i \mathbf{a}_i
\end{equation}
and get the machine-aware representation set $\mathbf{x_{g}}$. It is noted that the group representation set $\boldsymbol{G}$ is composed of $n_q$ trainable parameters of the network, which can represent specific characteristics of each group during training. And finally, a machine-aware representation $\mathbf{x_{g}}$ is added to the main branch to enhance the final representation with machine specific characteristics.

\begin{figure*}[h]
    \renewcommand{\thefigure}{4}
    \centering
    \includegraphics[width=1.0\linewidth]{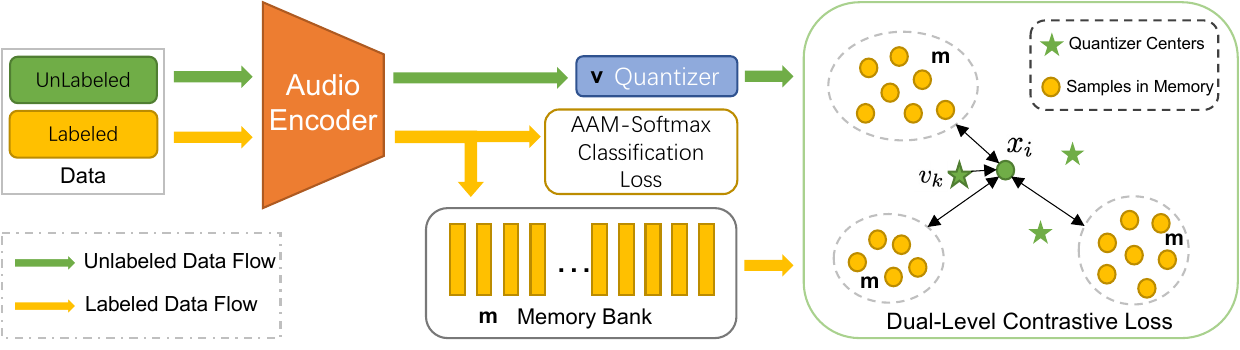}
    \caption{An overview of the proposed Dual-Level Contrastive Loss (DLCL), which can process the anomalous sound detection with partially attribute-unlabeled conditions. The \textcolor{DarkGreen}{green} line denotes the data lacking attributes information, and the \textcolor{orange}{orange} line represents the attributed data.}
    \label{fig:dlcl}
\end{figure*}

\subsection{Dual-Level Contrastive Loss for Model Optimization}
\label{sec:method_dlcl}
In our framework, the model uses discriminative approaches as an auxiliary loss for model optimization. To address the issue of missing attribute labels, we propose a new optimization objective based on contrastive learning.
An overview of the proposed dual-level contrastive loss is illustrated in Fig.~\ref{fig:dlcl}. All data are processed through a shared audio encoder (a pre-trained model in this work) to generate fixed-length audio representations, which are subsequently optimized in the embedding space. The optimization process comprises two main components: Annotated data are optimized using a classification loss, while unlabeled data are adjusted through dual-level contrastive learning.

For loss of classification, we apply Additive Angular Margin (AAM) softmax loss~\cite{arcface, aam} to machine audios by classifying metadata associated with each machine, a method which has been demonstrated to be effective for ASD in~\cite{wilkinghoff2023angular}. For the $i$-th sample and label $y_i$, it can be formulated as follows:
\begin{equation}
\label{equ:aam}
    L_{AAM} = -\frac{1}{N} \sum_{i=1}^{N} \log \frac{e^{s(\cos(\theta_{y_i,i} + m))}}{Z}
\end{equation}
where $Z=e^{s(\cos(\theta_{y_i,i} + m))} + \sum_{j=1, j\ne i}^c e^{s(\cos(\theta_{j,i}))}$, $\theta_{j, i}$ is the angle between the column vector $\mathbf{W_j}$ of classification head and input embedding $\mathbf{x_i}$, where both $\mathbf{W_j}$ and $\mathbf{x_i}$ are normalized. $s$ is a scaling factor and $m$ is a hyperparameter to control the margin. During classification optimization, embeddings derived from labeled data are also stored in a FIFO (first-in-first-out) memory bank $\mathbf{m}$, serving as negative samples for the subsequent contrastive learning phase.

To address the lack of attribute information in ASD, we employ contrastive objectives as additional supervision for unannotated data, implemented at both instance- and prototype-level. Suppose the current unattributed embedding is $\mathbf{x_i}$.
\begin{itemize}
    \item At the instance level, since unlabeled and labeled data originate from different machines, samples stored in the memory bank $\boldsymbol{M}=\{\mathbf{m_1}, \mathbf{m_2},\dots, \mathbf{m_{n_b}}\}$ are treated as negative samples, increasing the distance between the embedding $\boldsymbol{M}$ of memory bank and machine embedding $\mathbf{x_i}$. $n_m$ denotes the size of memory bank $\boldsymbol{M}$. 
    \item For prototype level, a vector quantizer-based online clustering method is adopted to generate dynamic pseudo-labels $\hat{y}_i$. The quantizer is composed of $n_q$ vectors, denoted as $\boldsymbol{V} = \{\mathbf{v_1}, \mathbf{v_2}, \dots, \mathbf{v_{n_q}}\}$. Pseudo label $\hat{y}_i$ of unlabeled sample $\mathbf{x_i}$ is generated by selecting the closest vector from the quantizer in terms of cosine similarity:
    \begin{equation}
    \label{equ:quantization}
        \hat{y}_i = \arg \min_{j\in [1,n_q]} \frac{\mathbf{x_i} \cdot \mathbf{v_j}}{|\mathbf{x_i}| \cdot |\mathbf{v_j}|}
    \end{equation}
    Suppose $\mathbf{v_c}$ is the closest vector of embedding $\mathbf{x_i}$. Then, we optimize the clustering process by minimizing the distance between this center vector $\mathbf{v_c}$ and the sample embedding $\mathbf{x_i}$. 
\end{itemize}

Therefore, the dual-level contrastive learning can be unified by (1) \textbf{increasing} the distance between the vectors from memory bank $\boldsymbol{M}$ and embedding $\mathbf{x_i}$ (2) \textbf{minimizing} the distance between the closest vector:
\begin{equation}
    L_{DLCL} = -\log \frac{\exp(\mathbf{x_i} \cdot \mathbf{v_c}/\tau_2)}{\exp(\mathbf{x_i} \cdot \mathbf{v_c}/\tau_2) + \sum_{j=1}^{n_b}\exp(\mathbf{x_i} \cdot \mathbf{m_j}/\tau_2)}
\end{equation}
where $\tau_2$ is the temperature factor. Then the final completed loss for the model optimization can be formulated as the sum of DLCL and AAM loss:
\begin{equation}
\label{equ:loss}
L = L_{AAM} + \lambda L_{DLCL}
\end{equation}
and $\lambda$ is a factor to balance the classification loss and the proposed DLCL Loss.

\section{Experiment Setup}

\subsection{Datasets}
For the sound anomaly detection task, the most widely used datasets are those from the DCASE challenge series~\cite{koizumi2020description, kawaguchi2021description, dohi2022description, dohi2023description, nishida2024description}. To evaluate the robustness and generalization of our proposed method, we conduct experiments on all datasets from the DCASE series, including five challenge datasets from 2020-2024, in contrast to the previous work that only evaluated performance on single datasets. This comprehensive evaluation can better show the superiority and generalization of the related methods.

The majority of data across these five datasets is generated using ToyADMOS~\cite{koizumi2019toyadmos}, ToyADMOS2~\cite{harada2021toyadmos2}, MIMII~\cite{purohit2019mimii, dohi2022mimii}, and IMAD-DS~\cite{Albertini2024}, all of which share a similar structure. Each dataset consists of three main components: (1) a development set, which includes a training subset with normal audio clips from various machines and a small validation subset, (2) an additional training set, and (3) an evaluation set containing both normal and anomalous audio clips. In addition to the audio clips, some datasets provide auxiliary metadata, including machine type, machine ID, and working condition attributes (It is noted that the DCASE2020 dataset does not include working condition information). The machine ID corresponds to distinct entities within each machine type.

In the DCASE 2022–2024 datasets, a clear distinction is made between source and target domains. The source samples comprise the majority (99\%) of the development and additional training data, representing known normal conditions. In contrast, the target samples correspond to the remaining 1\%, simulating unseen or rare conditions that are more challenging to detect.
In the evaluation set, the distributions of source and target samples are balanced and unknown to the system, making the detection task more realistic and challenging.

The annual datasets also exhibit slight variations to address different research challenges. For DCASE 2020~\cite{koizumi2020description}, both the development and evaluation datasets involve the same machine type, but do not include working condition attributes. In DCASE 2021~\cite{kawaguchi2021description} and DCASE 2022~\cite{dohi2022description}, the domain shift problem was introduced to build more robust ASD systems that account for variations in acoustic characteristics across different environments. In DCASE 2023~\cite{dohi2023description} and DCASE 2024~\cite{nishida2024description}, the focus shifted to evaluating the generalization of performance across different machine types, with the training and testing sets involving distinct machine types.

We will evaluate our method on all these datasets and aim to demonstrate its superiority across various aspects of performance.

\subsection{Evaluation Metrics}
For evaluation, the area under the receiver operating characteristic curve (AUC) was employed as a metric to assess the overall detection performance, while the partial-AUC (pAUC) was utilized to measure performance in a low false-positive rate (FPR) range $[0, p]$. The AUC and pAUC for each machine type are defined as following:
\begin{align}
    \mathrm{AUC} &= \frac{1}{N_{-} N_{+}} \sum_{i=1}^{N_{-}} \sum_{i=j}^{N_{+}} \mathcal{H}\left(\mathcal{A}_{\theta}\left(\boldsymbol{x}_{j}^{+}\right)-\mathcal{A}_{\theta}\left(\boldsymbol{x}_{i}^{-}\right)\right) \\
    p \mathrm{AUC} &=\frac{1}{\left\lfloor p N_{-}\right\rfloor N_{+}} \sum_{i=1}^{\left\lfloor p N_{-}\right\rfloor} \sum_{i=j}^{N_{+}} \mathcal{H}\left(\mathcal{A}_{\theta}\left(\boldsymbol{x}_{j}^{+}\right)-\mathcal{A}_{\theta}\left(\boldsymbol{x}_{i}^{-}\right)\right)
\end{align}
where $\left\lfloor \cdot \right\rfloor$ is the flooring function and $\mathcal{H}(a)$ is the hard-threshold function that returns 1 when $a > 0$ and 0 otherwise. Here, $\left\{\boldsymbol{x}_{i}^{-}\right\}_{i=1}^{N_{-}}$  and $\left\{\boldsymbol{x}_{j}^{+}\right\}_{j=1}^{N_{+}}$ are normal and anomalous test samples, respectively,
and have been sorted so that their anomaly scores are in descending order. Here, $N_{-}$ and $N_{+}$ are the numbers of normal and anomalous test samples, respectively. And $p$ is set to $0.1$ in this work.

\subsection{Baseline Systems}
\label{sec:baseline}
To provide a more comprehensive evaluation of our proposed framework, we compare it not only with publicly available state-of-the-art (SOTA) results and previous challenge-winning systems, but also with widely adopted baseline models.
For the common methods, we conducted code replication, including MobileNet~\cite{chen2018mobilefacenets}, which is a representative work and has been frequently used in previous challenge systems~\cite{Giri2020, LiuCQUPT2022, LopezIL2021}, demonstrating competitive performance. In addition, we incorporate systems based on AutoEncoder and classification methods, as provided by the official baseline~\cite{dohi2023description,harada2023first}, for comparison.

For most of the other comparison systems, the lack of publicly available implementation code, pre-trained model weights, and detailed training procedures makes it difficult for us to reproduce all methods independently.
As a result, the results reported in our tables are primarily collected from the original papers or official challenge reports, where only partial results are typically provided by the authors.
Nevertheless, we believe that even a partial comparison is sufficient to draw meaningful conclusions, as all systems are trained under the same DCASE configuration and dataset~\cite{koizumi2020description, kawaguchi2021description, dohi2022description, dohi2023description, nishida2024description}.
It is important to note that most top-tier systems from both challenges report ensemble models, whereas our system utilizes single models.

\subsection{Data Augmentation}
To generate additional training samples and enhance data diversity, we primarily employ SpecAug~\cite{specaug} as an online data augmentation strategy for audio pre-trained models. SpecAug applies a simple yet effective augmentation policy to the acoustic features, which includes frequency channel warping, time step masking, and frequency block masking. By modifying these aspects of the audio signal, SpecAug effectively simulates a variety of acoustic conditions, thus enriching the diversity of the training dataset.
For the speech pre-trained models that use raw waveform as input, SpecAug is not applied.

\subsection{Backend Detector}
For all systems developed in this work, we uniformly adopt the K-Nearest Neighbor (KNN) algorithm~\cite{ramaswamy2000efficient} as the backend method for detecting anomalous sounds. Specifically, the pre-trained audio encoder transforms normal audio samples into fixed-dimensional embeddings, which are then used to estimate the distribution of normal samples. For a new query audio, we compute the cosine distance between the query and its nearest neighbor (with $k=1$) and use this value as the anomaly score.
Moreover, different datasets exhibit distinct characteristics, prompting us to select different subsets for modeling the normal distribution. For the DCASE 2020 dataset, independent KNN detectors are employed for each machine type and machine ID. In contrast, for the DCASE 2021-2024 datasets, the division is based solely on machine type, as the evaluation sets do not provide machine ID labels. To address the issue of domain mismatch, we employ a soft scoring strategy~\cite{dohi2023description}, which initializes two KNN detectors—one for source samples and one for target samples. The anomaly score is then determined by selecting the lower of the two scores:
\begin{equation}
    S_{Anomaly} = \min(d_{source},d_{target})
\end{equation}
where $S$ is the anomalous score and $d$ represents the distance obtained by the corresponding detector.

\begin{table*}[th]
  \caption{Performance comparison of different speech and audio pre-trained models on \textbf{Development set of DCASE 2024}. It’s noted that all the results we report are the harmonic mean (hmean) of the AUC and pAUC following~\cite{nishida2024description}. AUC\_s and AUC\_t denote the AUC of source and target samples, respectively.}
  \label{tab:dcase24_pretrain_dev}
  \centering
  \scalebox{0.93}{
  \begin{tabular}{cccc|ccccccc|ccc}
    \toprule
    \multirow{2}*{\textbf{Set}} & \multirow{2}*{\textbf{Domain}} & \multirow{2}*{\textbf{Models}} & \multirow{2}*{\textbf{Size}} & \multicolumn{7}{c|}{\textbf{Machines (Hmean)}} & \multicolumn{3}{c} {\textbf{All}} \\
     & & & & Bearing & Fan & Gearbox & Slider & ToyCar & ToyTrain & Valve & AUC\_s & AUC\_t & \textbf{Hmean} \\
    \midrule
    \multirow{13}*{\bf Dev} & \multirow{3}*{-} & AutoEncoder~\cite{nishida2024description,harada2023first} & - & 49.57 &	54.86 &	59.64 &	59.50 &	\textbf{62.76} &56.52 &	50.54 & 65.15 & 51.23 & 55.83 \\ 
    & & CNN10~\cite{koizumi2020description} & - & 63.66 & 62.31 & 65.10 & 59.47 & 54.50 & 55.39 & 68.58 & 64.47 & 63.63 & 60.46 \\ 
    & & MobileNet~\cite{morita2021anomalous} & - & 63.33 &	\textbf{64.55} &	59.58 &	69.04 &	54.11 &	49.84 &	53.66 & 62.56 & 59.62 & 58.40 \\ 
    \cmidrule{2-14}
    & \multirow{6}*{Speech} & Wav2Vec 2.0~\cite{baevski2020wav2vec} & 316M & 65.19 & 63.26 &\textbf{70.10} & 51.23 &54.79 & 52.48 &	59.71 & 62.02 & 59.73 &	58.78 \\
    & & UniSpeech~\cite{wang2021unispeech} & 316M & 58.43 & 57.57 & 64.93 &51.74 &	51.82 &	53.45 &	63.66 & 58.37 & 58.96 &	56.86 \\
    & & HuBERT-Base~\cite{hsu2021hubert} & 95M & 62.45 &	56.05 &	68.80 &	52.15 &	48.36 &	51.21 &	64.91 & 56.06 & 60.43 &	56.75 \\
    & & HuBERT-Large~\cite{hsu2021hubert} & 316M & 60.49 &	57.91 &	60.65 &	51.85 &	49.08 &	51.58 &	68.66 & 58.65 & 57.82 &	56.42  \\
    & & WavLM-Base~\cite{chen2022wavlm} & 95M & 61.00 &	56.22 &	67.03 &	49.85 &	49.49 &	55.44 &	61.61 & 57.17 & 59.94 &	56.58 \\
    & & WavLM-Large~\cite{chen2022wavlm} & 316M & 53.66 &	59.61 &	67.66 &	47.49 &	51.67 &	54.06 &	60.65 & 57.37 & 55.63 &	55.76 \\
    \cmidrule{2-14}
    & \multirow{6}*{Audio} & AST~\cite{gong21b_interspeech} & 86M & 57.08 &	63.15 &	61.13 &	\textbf{72.60} & 50.45 & 56.92 & 62.85 & 60.12 & 63.39 & 59.87  \\
    & & ATST~\cite{li2024self} & 85M & \textbf{67.76} &	60.26 &	64.28 &	57.12 &	51.37 &	56.05 &	69.87 & 63.07 & 63.60 & 60.25 \\
    & & CED ~\cite{ced} & 85M & 61.80 & 62.53 & 63.36 & 57.12 & 53.30 & 65.73 & 61.94 & 64.32 & 63.63 & 60.53 \\
    & & CLAP-LAION ~\cite{clap-laion} & 194M & 63.54 & 61.69 & 59.86 & 56.95 & 51.94 & 54.21 & 62.73 & 58.53 & 62.86 & 58.29 \\
    & & Imagebind~\cite{girdhar2023imagebind} & 86M & 65.14 &	61.05 &	64.42 &	69.71 &	54.50 &	51.25 &	66.10 & 65.81 & 64.06 &	60.76 \\
    & & BEATs~\cite{beats} & 90M & 65.38 & 62.04 & 66.38 & 62.27 & 53.40 & \textbf{59.50} & \textbf{71.09} & \textbf{67.77} & \textbf{67.74} & \textbf{62.42} \\
    \bottomrule
  \end{tabular}
  }
\end{table*}

\begin{table*}[th]
  \caption{Performance comparison of different speech and audio pre-trained models on the \textbf{Evaluation set of DCASE 2024}. It’s noted that all the results we report are the harmonic mean (hmean) of the AUC and pAUC following~\cite{nishida2024description}. The machine types of the evaluation set and the training set are different. Due to space limitations, machine types are abbreviated. AUC\_s and AUC\_t denote the AUC of source and target samples, respectively.}
  \label{tab:dcase24_pretrain_eval}
  \centering
  \scalebox{0.85}{
  \begin{tabular}{cccc|ccccccccc|ccc}
    \toprule
    \multirow{2}*{\textbf{Set}} & \multirow{2}*{\textbf{Domain}} & \multirow{2}*{\textbf{Models}} & \multirow{2}*{\textbf{Size}} & \multicolumn{9}{c|}{\textbf{Machines (Hmean)}} & \multicolumn{3}{c} {\textbf{All}} \\
& & & & 3DPrin. & A.Com. & B.Mot. & H.Dry. & H.Dro. & R.Arm & Scan. & T.Bru. & T.Cir. & AUC\_s & AUC\_t & \textbf{Hmean} \\
    \midrule
    \multirow{13}*{\textbf{Eval}} & \multirow{3}*{-} & AutoEncoder~\cite{nishida2024description,harada2023first} & - & 54.57 & 55.43 & 61.25 & 52.26 & 54.16 & 51.06 & 55.30 & 62.45 & 56.21 & \textbf{71.51} & 50.58 & 55.63 \\
    & & CNN10~\cite{koizumi2020description} & - & 57.60 & 55.85 & 56.58 & 57.04 & 60.00 & 54.95 & 57.80 & 53.98 & 58.30 & 60.89 & 57.43 & 56.73\\ 
    & & MobileNet~\cite{morita2021anomalous} & - & 57.43 & 56.61 & 56.94 & 55.45 & 56.25 & 54.05 &	52.10 &	54.60 &	55.85 & 56.76 & 57.84 &	55.28  \\ 
    \cmidrule{2-16}
    & \multirow{6}*{Speech} & Wav2Vec 2.0~\cite{baevski2020wav2vec} & 316M & 53.07 & 56.57 &	\textbf{67.64} &	52.89 &	55.17 &	55.01 &	69.85 & 62.79 	& 61.07 & 62.25 & 61.26 &	58.70 \\
    & & UniSpeech~\cite{wang2021unispeech} & 316M & 55.03 &	49.13 &	58.66 &	49.00 &	54.24 &	58.50 &	71.96 & 	59.94 &	64.69 & 61.26 & 57.72 &	56.92 \\
    & & HuBERT-Base~\cite{hsu2021hubert} & 95M & 56.20 &	49.29 &	57.22 &	48.84 &	55.92 &	56.59 &	65.49 &	64.59 &	59.52 & 59.41 & 58.63 &	56.47 \\
    & & HuBERT-Large~\cite{hsu2021hubert} & 316M & 52.78 &	50.48 &	60.84 &	47.43 & 51.33 &	52.36 &	80.37 &	60.22 &	62.85 & 60.15 & 56.37 &	56.26 \\
    & & WavLM-Base~\cite{chen2022wavlm} & 95M & 52.95 & 50.85 &	54.21 &	50.79 &	53.87 & 53.79 & 70.09 & 59.13 & 60.80 & 59.58 & 55.80 &	55.61 \\
    & & WavLM-Large~\cite{chen2022wavlm} & 316M & \textbf{58.93} &	51.00 &	58.78 &	52.51 &	56.65 & 55.40 &	79.43 &	59.25 &	64.33 & 61.28 & 57.82 &	58.60\\
    \cmidrule{2-16}
    & \multirow{6}*{Audio} & AST~\cite{gong21b_interspeech} & 86M & 53.12 &	52.55 &	51.89 &	47.89 &	58.09 &	56.76 &	57.76 &	55.91 &	55.67 & 57.58 & 53.17 &	54.07  \\
    & & ATST~\cite{li2024self} & 85M & 54.15 &	51.99 & 67.03 & 47.73 & 58.45 &  59.38 &	65.07 &	58.55 &	56.00 & 59.73 & 59.41 &	56.88 \\
    & & CED~\cite{ced} & 85M & 60.42 & 50.87 & 58.65 & 56.43 & 53.07 & 68.09 & 92.55 &	66.22 & 62.59 & 63.46 & 65.68 & 60.96 \\
    & & CLAP-LAION~\cite{clap-laion} & 194M & 54.03 & 51.83 & 53.38 & 49.09 & 57.19 & 59.32 &	69.88 &	57.89 &	58.07 & 57.08 & 58.56 & 57.11 \\
    & & Imagebind~\cite{girdhar2023imagebind} & 86M & 56.67 &	\textbf{61.96} &	66.38 &	58.24 &	60.96 &	54.88 &	57.35 &	58.93 &	61.36 & 63.62 & 61.71 &	59.38 \\
    & & BEATs~\cite{beats} & 90M & 56.74 & 55.99 & 59.91 & \textbf{66.59} & \textbf{67.35} & \textbf{64.58} & \textbf{85.44} & \textbf{64.92} & \textbf{66.66} & 68.69 & \textbf{71.25} & \textbf{64.46} \\
    \bottomrule
  \end{tabular}
}
\end{table*}

\subsection{Training Configuration}

For the baseline model, we utilize the original configuration provided by the official implementation during training. Regarding the fine-tuning process in this work, there are slight variations in configuration depending on the specific pre-trained model. For speech models, such as Wav2Vec 2.0 and HuBERT, the waveform is used as input directly. In contrast, for audio models, 128-dimensional Mel-Frequency Cepstral Coefficients (MFCCs) are employed as input acoustic features, which are computed using a 25 ms window length, 10 ms hop length, and a Hamming window.

During training, each audio file is segmented into 2-second chunks, which are then fed into the models. The learning rate scheduler follows a warm-up strategy for the first 960 steps, starting with an initial learning rate of 0.0001. The AdamW optimizer is used to update the model parameters. A batch size of 32 is applied, and the fine-tuning process continues for a total of 60,000 steps until convergence. In addition, when applying group adapter and DLCL, hyper-parameters $\tau_1$, $\tau_2$, $n_q$, and $\lambda$ are set to 0.1, 0.1, 64, and 1.0, respectively.

\section{Results and Analysis}

The experiments are divided into five sections. In Section~\ref{sec:res_pretrain}, we present a performance comparison of various audio and speech pre-trained models on the anomaly sound detection task. We also analyze whether the observed performance improvements stem from the model parameters or the pretraining process itself. In Section~\ref{sec:res_lora}, we report the results of applying fully-connected multi-branch LoRA during the fine-tuning process and provide an ablation study on the effects of rank and branch number in the fully-connected layer (FCL). Section~\ref{sec:res_groupadapter} investigates the impact of the position and number of adapters added to the Group Adapter. Section~\ref{sec:res_loss} examines the effectiveness of the newly proposed DLCL loss function in the absence of machine operating condition attributes. Finally, Section~\ref{sec:res_overview} offers a more comprehensive comparison between our newly proposed system and previous works using the DCASE 2020-2024 five challenge datasets.

\subsection{Evaluation of Pre-trained Models for ASD}
\label{sec:res_pretrain}
Table~\ref{tab:dcase24_pretrain_dev} and Table~\ref{tab:dcase24_pretrain_eval} present the performance comparison of fine-tuning various pre-trained models on the ASD task, along with other baseline systems. All models are trained on the DCASE 2024 dataset and evaluated on the development and evaluation sets, respectively.

In Table~\ref{tab:dcase24_pretrain_dev}, we first examine the speech-based pre-trained models, including Wav2Vec~\cite{baevski2020wav2vec}, UniSpeech~\cite{wang2021unispeech}, Hubert~\cite{hsu2021hubert}, and WavLM~\cite{chen2022wavlm}. Originally designed for automatic speech recognition (ASR), these models are pre-trained on large-scale speech datasets to extract frame-level semantic information. Despite employing different self-supervised learning approaches, these models show limited performance gains when fine-tuned for anomaly sound detection. Additionally, when comparing different sizes of speech pre-trained models, larger models, which are trained on more extensive speech datasets, do not demonstrate any performance improvement for ASD. This suggests a discrepancy between the data characteristics and objectives of speech pre-trained models and the requirements of the ASD task.

Next, we analyze several representative audio-based pre-trained models, including AST~\cite{gong21b_interspeech}, ATST~\cite{li2024self}, CED~\cite{ced}, CLAP-LAION~\cite{clap-laion}, ImageBind~\cite{girdhar2023imagebind}, and BEATs~\cite{beats}. 
They all were designed for audio understanding and pre-trained on large-scale audio datasets~\cite{gemmeke2017audio} to capture patch-level acoustic information. However, these models differ significantly in their training strategies.
Models such as AST, ATST, and CED are pre-trained on AudioSet using supervised learning with labeled classification objectives. These models are typically fine-tuned for downstream audio classification tasks.
In contrast, ImageBind and CLAP-LAION adopt a contrastive learning paradigm and utilize paired multimodal data: ImageBind uses audio–text–image triplets, while CLAP employs audio–text pairs. These models are primarily optimized for cross-modal representation alignment and have achieved leading performance in cross-modal tasks.
BEATs, on the other hand, is trained without any labels. It relies solely on self-supervised learning via a mask-and-predict strategy to learn powerful audio representations, which are then fine-tuned for various downstream tasks. By comparing the performance of various types of audio pre-training models, we observe that the self-supervised BEATs model~\cite{beats} tends to outperform other supervised pre-trained models in anomalous sound detection tasks.
This suggests that the representations learned through BEATs’ self-supervised training, which are driven by the intrinsic structure and correlations within unlabeled sound data, are capable of capturing fundamental, task-agnostic features with stronger transferability across the anomalous sound detection task.
In contrast, representations learned through supervised pre-training are often task-specific, optimized toward the labeled objectives present in the training data, and thus less adaptable to unseen tasks or domains—particularly since anomalous machine sounds are virtually absent from the pre-training datasets.

Compared to speech-based models, these audio-based models, even having fewer parameters, still significantly outperform speech models on the development set of DCASE 2024. This indicates that audio-based pre-training aligns more closely with the requirements for modeling machine operation sounds in the ASD task. To facilitate the comparison, we selected several commonly used baseline systems, as described in Section~\ref{sec:baseline}. Among these, Autoencoder and CNN10 are the officially provided baseline systems~\cite{dohi2023description}; Autoencoder is optimized through spectrogram reconstruction, while CNN10 is based on attribute classification, similar to MobileNet~\cite{morita2021anomalous}. Analysis of the results shows that classification-based models outperform the reconstruction-based autoencoder model. Our pre-trained models are optimized with classification objectives, demonstrating significant advantages over these baseline systems.

Then we present the performance comparison of various models on the DCASE 2024 Evaluation set in Table~\ref{tab:dcase24_pretrain_eval}, which covers the unseen machine types of the training set. It is observed that BEATs model~\cite{beats} achieves the best performance, and its advantage and improvement are even larger and more obvious on the Evaluation set compared to all the other models.

\begin{table}[h]
    \centering
    \caption{Performance comparison on the DCASE 2024 dataset of initializing models with or without the pre-trained weights. All hmean is the harmonic mean of both the Development and Evaluation sets.}
    \begin{tabular}{cc|cc|c}
    \toprule
       \multirow{2}*{\textbf{Models}}  & \multirow{2}*{\textbf{Pretrained}} & \multicolumn{3}{c}{\textbf{Hmean}}\\
       & & Dev & Eval & All \\ \midrule
       AutoEncoder~\cite{nishida2024description} & \usym{2717} & 56.20 & 55.85 & 56.00 \\ 
       CNN10~\cite{koizumi2020description} & \usym{2717} & 60.46 & 56.43 & 58.53 \\ 
       MobileNet~\cite{morita2021anomalous} & \usym{2717} & 58.40 & 55.28 & 56.79 \\ \midrule
       Wav2Vec 2.0~\cite{baevski2020wav2vec} & \usym{2717} & 55.23 & 55.53 & 55.38 \\
       Wav2Vec 2.0~\cite{baevski2020wav2vec} & \checkmark & 58.78 & 58.70 & 58.73 \\
       BEATs~\cite{beats}  & \usym{2717} & 60.05 & 57.69 & 58.84 \\ 
       BEATs~\cite{beats}  & \checkmark & 62.42 & 64.46 & 63.42 \\
       \bottomrule
    \end{tabular}
    \label{tab:train_from_scratch}
\end{table}

To explore the source of these performance improvements, we conducted an ablation study to examine the effect of initializing models with or without the pre-trained weights. We selected wav2vec 2.0~\cite{baevski2020wav2vec} and BEATs~\cite{beats} as representative pre-trained models for speech and audio, respectively, with results presented in Table~\ref{tab:train_from_scratch}. For both models, a random initialization (keeping the model structure unchanged) leads to a substantial performance decline on both the development and evaluation sets, even worse than the baseline models. This indicates that the performance gains stem from the knowledge embedded in the pre-training process, rather than merely from the larger parameter size or the transformer-based architecture of the pre-trained model.
Although a potential mismatch between pre-training data and downstream tasks, the pre-training process can provide a strong initialization for fine-tuning robust machine sound encoders.

In summary, Tables \ref{tab:dcase24_pretrain_dev},\ref{tab:dcase24_pretrain_eval}, and \ref{tab:train_from_scratch} present the performance of speech and audio pre-trained models on the sound anomaly detection task. Specifically, for audio pre-trained models, we explored models derived through various training strategies, including supervised training, multi-modal training, and self-supervised training. Through this exploration, we found that BEATs~\cite{beats}, a model pre-trained via self-supervised learning on a large-scale audio dataset, exhibits excellent generalization ability. Consequently, we adopt it as the default backbone in our proposed framework and employ it in the subsequent experiments.

\subsection{Evaluation of Fully Connected Multi-Branch LoRA}
\label{sec:res_lora}

\begin{figure}[t]
    \centering
    \begin{subfigure}{0.5\textwidth}
        \centering
        \includegraphics[width=0.9\linewidth]{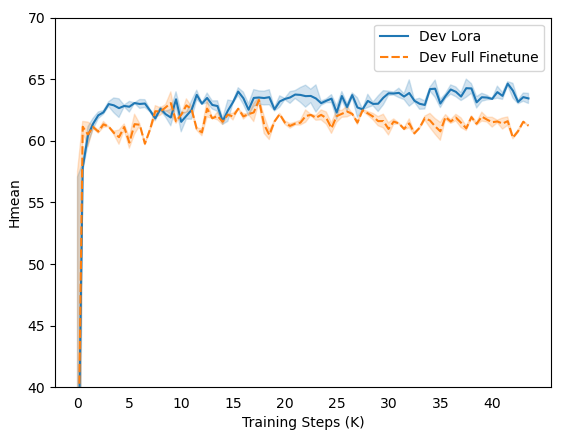}
        \caption{Hmean across different steps on Dev set}
    \end{subfigure}
    \begin{subfigure}{0.5\textwidth}
        \centering
        \includegraphics[width=0.9\linewidth]{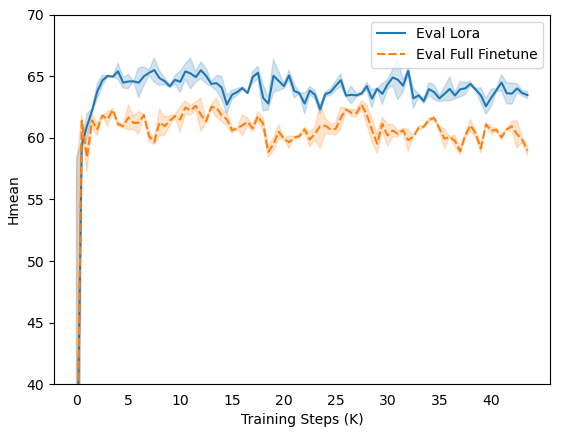}
        \caption{Hmean across different steps on Eval set}
    \end{subfigure}
    \caption{Performance comparison visualization on DCASE 2024 dataset for the systems with or without the proposed fully connected multi-branch LoRA. The above figure(a) shows the results on the Development set, and the below figure(b) shows the results on the Evaluation set. The machine types in the Evaluation set are different from those in the Development set.}
    \label{fig:lora_robust}
\end{figure}
\begin{table}[h]
    \centering
    \caption{Performance comparison of fine-tuning fully-connected multi-branch low-rank adapter module with different ranks and branch number on DCASE 2024 dataset. All hmean is the harmonic mean of both the Development and Evaluation sets. Result with underline ``\underline{~~}'' means that LoRA branch number equals 1, which is LoRA baseline~\cite{hu2021lora}. FC denotes that multi-branch LoRA with the fully-connected strategy.}
    \begin{tabular}{cccc|cc|c}
    \toprule
    \multirow{2}*{\textbf{Rank}} & \textbf{LoRA} & \multirow{2}*{\textbf{Frozen}} & \multirow{2}*{\textbf{FC}} & \multicolumn{3}{c}{\textbf{Hmean}} \\
    & \textbf{Num.} & & & Dev & Eval & All \\ \midrule
    Full Finetune & - & - & - & 62.42 & 64.46 & 63.42 \\ 
    \midrule
    \cellcolor[HTML]{EFEFEF}8 & 8 & \checkmark & \checkmark & 64.00 & 62.61 & 63.30 \\
    \cellcolor[HTML]{EFEFEF}16 & 8 & \checkmark & \checkmark & 63.36 & 64.87 & 64.11 \\
    \cellcolor[HTML]{EFEFEF}64 & 8 & \checkmark & \checkmark & \textbf{64.22} & 64.49 & 64.35 \\
    32 & \cellcolor[HTML]{EFEFEF}1 & \checkmark & \checkmark & \underline{61.79} & \underline{64.31} & \underline{63.02} \\
    32 & \cellcolor[HTML]{EFEFEF}4 & \checkmark & \checkmark & 62.65 & 65.43 & 64.01 \\
    32 & \cellcolor[HTML]{EFEFEF}16 & \checkmark & \checkmark & 62.11 & 63.72 & 62.90 \\
    32 & 8 & \cellcolor[HTML]{EFEFEF}\usym{2717} & \checkmark & 62.56 & 66.33 & 64.39 \\
    32 & 8 & \checkmark & \cellcolor[HTML]{EFEFEF}\usym{2717} & 63.25 & 65.82 & 64.67 \\
    \midrule
    32 & 8 & \checkmark & \checkmark & 63.67 & \textbf{66.69} & \textbf{65.15} \\
    \bottomrule
    \end{tabular}
    \label{tab:lora}
\end{table}

To evaluate the performance of our proposed fully connected multi-branch LoRA under different hyperparameters, we conduct an ablation study on factors such as rank and the number of LoRA branches, with the results presented in Table~\ref{tab:lora}. Compared to the baseline with full fine-tuning, freezing the original weights and fine-tuning only the corresponding LoRA modules yields better performance on the DCASE 2024 dataset in most cases. 
When the number of branches is equal to 1, it represents a single-branch LoRA, which is the baseline of the base LoRA, but exhibits poorer performance than full fine-tuning. After freezing the parameters, only fine-tuning the single branch LoRA lacks the ability to fit to ASD tasks.
The optimal performance is achieved when the rank is set to 32 and the number of LoRA branches is set to 8. 
In contrast, when the original weights are unfrozen and optimized via gradient descent during fine-tuning, the performance declines, suggesting that an increased number of trainable parameters leads to overfitting and detrimental effects. In addition, ablation experiments were conducted on fully connected (FC) layers among multiple branches, demonstrating that connections between multiple branches have a positive impact on performance.

To further assess the effectiveness of our approach, the performance as a function of training steps is shown in Fig.~\ref{fig:lora_robust}. The figure reveals that, compared to the full fine-tuning, the proposed fully connected multi-branch LoRA obtains significant improvements on both the DCASE 2024 Development and Evaluation sets. Moreover, the improvements on the Evaluation sets are larger than those on the Development sets. The reason is that the Evaluation set is evaluated on different machine types, which are unseen in the training set, while the machine types in the Development set are the same as those in the training set. This indicates that the proposed fully connected multi-branch LoRA helps prevent the forgetting of pre-trained knowledge by freezing the original weights while fine-tuning with a small number of parameters, thereby mitigating overfitting and constructing a more generalized ASD model.

\subsection{Evaluation of Machine-aware Group Adapter}
\label{sec:res_groupadapter}
Similarly, we conduct ablation experiments on the Machine-aware Group Adapter, with results presented in Table~\ref{tab:adapter}. First, with regard to the placement of the Group Adapter, we find that inserting it into the multi-head attention (MHA) layer yields better performance than placing it in the entire transformer layer or the feed-forward network (FFN) module. Furthermore, our exploration of the optimal group number reveals that 32 is the most effective configuration.

\begin{table}[h]
    \centering
    \caption{Performance comparison of the proposed Group Adapter with different group numbers on the DCASE 2024 dataset. All hmean is the harmonic mean of both the Development and Evaluation sets. FFN and MHA denote feed-forward network and multi-head attention blocks, respectively.}
    \begin{tabular}{cc|cc|c}
    \toprule
    \textbf{Pos.} & \textbf{Group Num.} & \textbf{Dev} & \textbf{Eval} & \textbf{Hmean} \\ \midrule
    w/o Adapter & - & 62.42 & 64.46 & 63.42  \\ \midrule
    \cellcolor[HTML]{EFEFEF}Transformer & 32 & 64.15 & \textbf{65.61} & 64.87 \\
    \cellcolor[HTML]{EFEFEF}FFN & 32 & 63.90 & 64.59 & 64.24 \\
    MHA & \cellcolor[HTML]{EFEFEF}8 & 64.24 & 64.71 & 64.47 \\
    MHA & \cellcolor[HTML]{EFEFEF}16 & 64.24 & 65.40 & 64.81 \\
    MHA & \cellcolor[HTML]{EFEFEF}64 & \textbf{65.21} & 64.65 & 64.93 \\ \midrule
    MHA & 32 & 65.04 & 65.41 & \textbf{65.22} \\ 
    \bottomrule
    \end{tabular}    
    \label{tab:adapter}
\end{table}

\begin{figure}[h]
    \centering
    \includegraphics[width=1.0\linewidth]{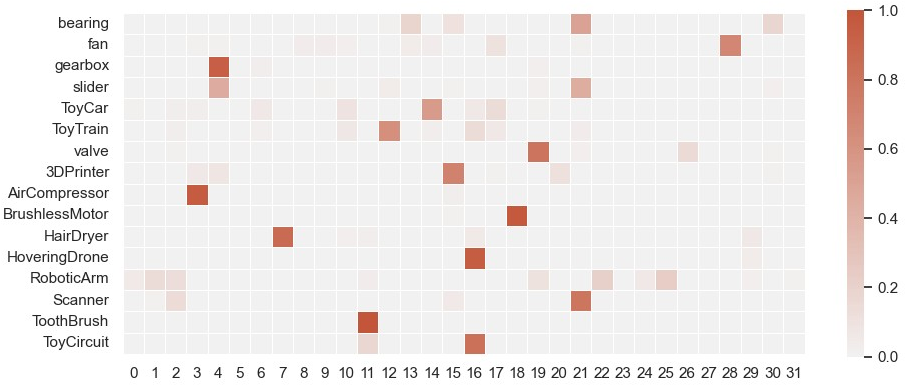}
    \caption{The visualization of Machine-aware Group Adapter normalized weights versus different machine types in the evaluation set of DCASE 2024. The darker the color, the greater the weight.}
    \label{fig:adapter}
\end{figure}

We also visualized the normalized weights of the Group Adapters for the different machine types in the Evaluation set, as shown in Fig.~\ref{fig:adapter}. Although the machine types in the evaluation and training are different, each audio sample can still be automatically clustered into distinct groups, demonstrating that the model can adapt its representations to the specific characteristics of different machines. This capability explains why the proposed Machine-aware Group Adapter can effectively improve the model performance.

\begin{figure*}
    \begin{subfigure}{0.5\textwidth}
        \centering
        \includegraphics[width=0.9\linewidth]{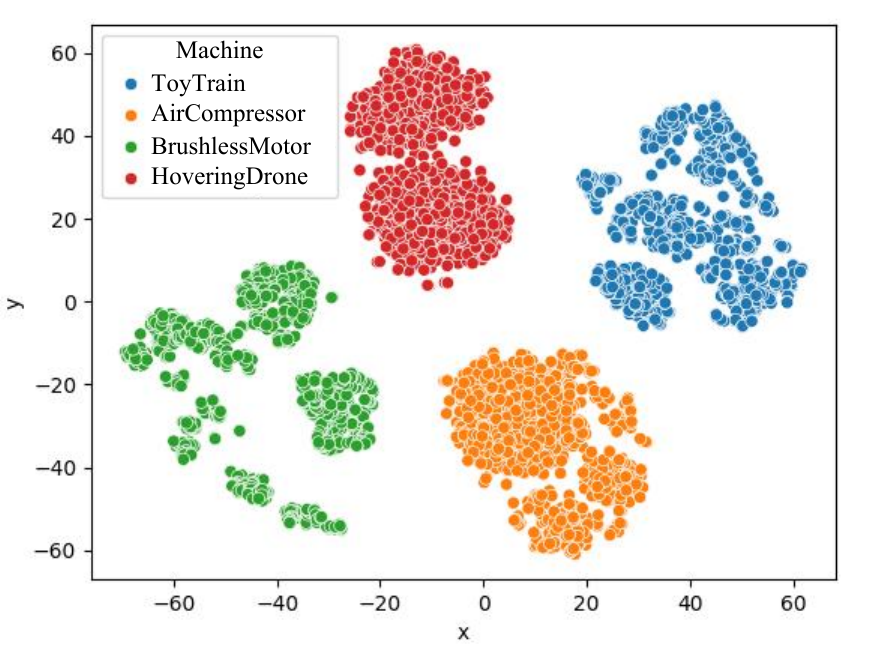}
        \caption{AAM-Softmax on Unlabeled Data}
    \end{subfigure}
    \begin{subfigure}{0.5\textwidth}
        \centering
        \includegraphics[width=0.9\linewidth]{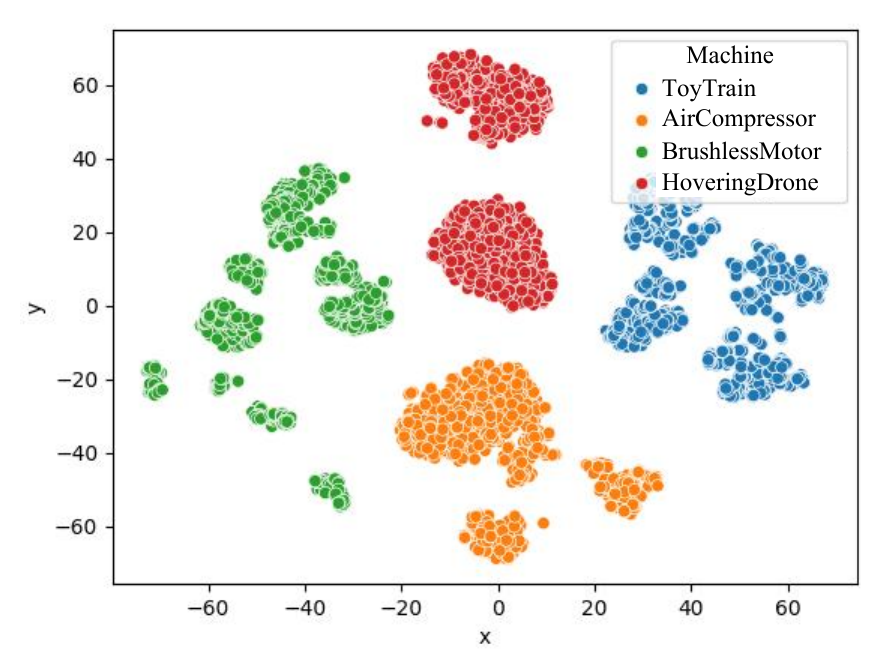}
        \caption{DLCL Loss on Unlabeled Data}
    \end{subfigure}
    \begin{subfigure}{0.5\textwidth}
        \centering
        \includegraphics[width=0.9\linewidth]{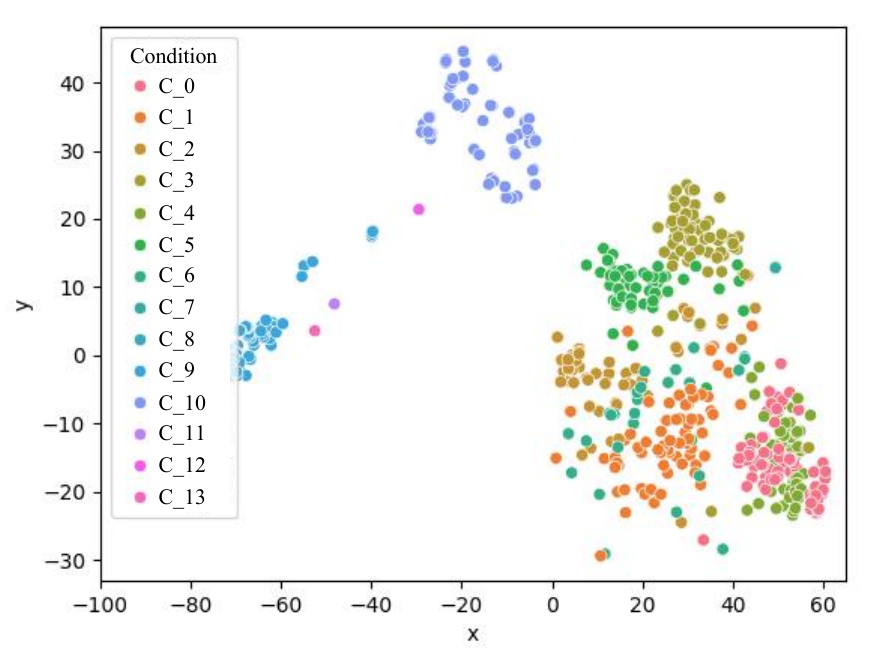}
        \caption{AAM-Softmax on Labeled Data}
    \end{subfigure}
    \begin{subfigure}{0.5\textwidth}
        \centering
        \includegraphics[width=0.9\linewidth]{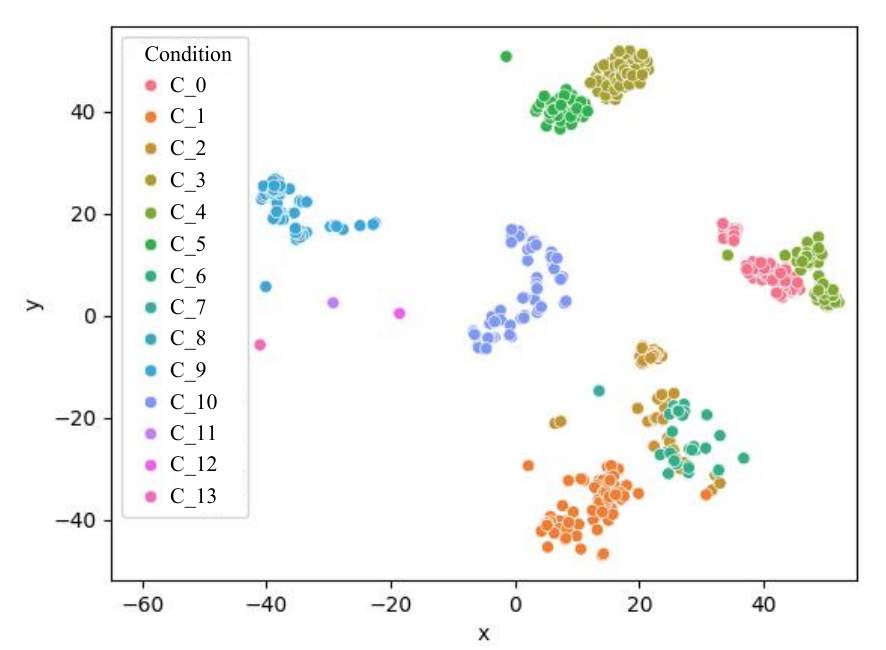}
        \caption{DLCL Loss on Labeled Data}
    \end{subfigure}
    \caption{T-SNE visualization of machines with missing attributes from the DCASE 2024 dataset. The subfigure (a) is the model fine-tuned with only AAM-Softmax classification loss. And the subfigure (b) is the model fine-tuned with our proposed dual-level contrastive learning (DLCL) loss during the training process. In the two subfigures (c) and (d), machines HairDryer and RoboticArm are trained without attribute labels, but visualized with ground truth attributes to illustrate the effects of two different losses.}
    \label{fig:res_loss}
\end{figure*}

\begin{table*}[h]
  \caption{Performance comparison on the \textbf{DCASE 2020 dataset} between our proposed model and previous works. The pre-trained model is BEATs~\cite{beats}. Both fully-connected multi-branch LoRA and group adapter are applied for fine-tuning. All the results are reported using the mean of the AUC and pAUC following~\cite{koizumi2020description}.}
  \label{tab:dcase20}
  \centering
  \adjustbox{width=1.0\linewidth,center=\linewidth}{
  \begin{tabular}{c|ccccccc|ccccccc|c}
    \toprule
    \multirow{2}*{Models} & \multicolumn{7}{c|}{\textbf{Development set}} & \multicolumn{7}{c|}{\textbf{Eval set}} & \textbf{All} \\
     & Fan & Pump & Slider & T.Car & T.Conv. & Valve & \textbf{mean} & Fan & Pump & Slider & T.Car & T.Conv. & Valve & \textbf{mean} & \textbf{mean} \\
     \midrule
     \color[HTML]{9B9B9B}2020 No.1~\cite{Giri2020} & \color[HTML]{9B9B9B}80.65 & \color[HTML]{9B9B9B}83.27 & \color[HTML]{9B9B9B}93.41 & \color[HTML]{9B9B9B}92.72 & \color[HTML]{9B9B9B}73.28 & \color[HTML]{9B9B9B}94.30 & \color[HTML]{9B9B9B}86.27 & \color[HTML]{9B9B9B}89.42 & \color[HTML]{9B9B9B}87.69 & \color[HTML]{9B9B9B}93.68 & \color[HTML]{9B9B9B}92.04 & \color[HTML]{9B9B9B}82.27 & \color[HTML]{9B9B9B}93.51 & \color[HTML]{9B9B9B}89.77 & \color[HTML]{9B9B9B}88.02 \\
     \midrule
     IDNN~\cite{suefusa2020anomalous} & 60.31 & 67.42 & 77.02 & 73.96 & 65.39 & 74.52 & 69.77 & - & - & - & - & - & - & - & - \\ 
     SCAdaCos~\cite{wilkinghoff2021sub} & 82.77 & 91.81 & 98.59 & 94.01 & 76.77 & 96.63 & 90.10 & 95.42 & 92.53 & 93.54 & 93.96 & 84.94 & 97.31 & 92.95 & 91.53 \\
     MFN~\cite{Giri2020} & 83.71 & 90.82 & 98.70 & 91.97 & 71.29 & 96.49 & 88.83 & 94.72 & 92.94 & 97.58 & 94.31 & 77.54 & 94.88 & 92.00 & 90.38 \\
     STgram~\cite{liu2022anomalous} & 91.51 & 86.85 & 98.58 & 91.06 & 69.09 & 99.04 & 89.35 & - & - & - & - & - & - & - & - \\
     Glow aff~\cite{dohi2021flow} & 70.10 & 78.60 & 88.70 & 88.15 & 65.25 & 83.20 & 79.55  & - & - & - & - & - & - & - & - \\
     SWNET~\cite{chen2023sw} & 94.54 & 84.98 & 96.77 & 92.85 & 74.70 & 98.14 & 90.33 & - & - & - & - & - & - & - & - \\
     PAE~\cite{zeng2023joint} & 72.53 & 69.48 & 63.66 & 68.14 & 83.16 & 88.33 & 74.22 & - & - & - & - & - & - & - & -  \\
     GeCo~\cite{zeng2023joint} & 88.80 & 90.32 & 97.75 & 94.40 & 74.32 & 98.34 & 90.65 & - & - & - & - & - & - & - & -  \\
     ASD-AFPA~\cite{zhang23fa_interspeech} & 95.51 & 90.61 & 99.04 & 97.27 & 92.80 & 70.35 & 90.93 & - & - & - & - & - & - & - & -  \\
     Ours & 85.85 & 94.03 & 98.11 & 97.12 & 74.72 & 96.68 & \textbf{91.08} & 95.1 & 96.11 & 99.49 & 97.92 & 84.26 & 97.17 & \textbf{95.01} & \textbf{93.05} \\
    \bottomrule
  \end{tabular}
  }
\end{table*}

\begin{table*}[h]
  \caption{Performance comparison on the \textbf{DCASE 2021 dataset} between our proposed model and previous works. The pre-trained model is BEATs~\cite{beats}. Both fully-connected multi-branch LoRA and group adapter are applied for fine-tuning. All the results are reported using the harmonic mean of the AUC and pAUC across different machines following~\cite{kawaguchi2021description}.}
  \label{tab:dcase21}
  \centering
  \adjustbox{width=1.0\linewidth,center=\linewidth}{
  \begin{tabular}{c|cccccccc|cccccccc|c}
    \toprule
    \multirow{2}*{Models} & \multicolumn{8}{c|}{\textbf{Development set}} & \multicolumn{8}{c|}{\textbf{Eval set}} & \textbf{All} \\
     & T.Car & T.Tra & Fan & G.box & Pump  & Slider & Valve & \textbf{Hmean} & T.Car & T.Tra & Fan & G.box & Pump  & Slider & Valve & \textbf{Hmean} & \textbf{Hmean} \\
     \midrule
     \color[HTML]{9B9B9B}2021 No.1~\cite{LopezIL2021} & \color[HTML]{9B9B9B}82.66 & \color[HTML]{9B9B9B}73.85 & \color[HTML]{9B9B9B}80.73 & \color[HTML]{9B9B9B}80.95 & \color[HTML]{9B9B9B}73.28 & \color[HTML]{9B9B9B}75.45 & \color[HTML]{9B9B9B}83.27 & \color[HTML]{9B9B9B}78.39 & \color[HTML]{9B9B9B}66.59 & \color[HTML]{9B9B9B}64.20 & \color[HTML]{9B9B9B}60.90 & \color[HTML]{9B9B9B}62.31 & \color[HTML]{9B9B9B}84.07 & \color[HTML]{9B9B9B}72.08 & \color[HTML]{9B9B9B}62.65 & \color[HTML]{9B9B9B}66.79 & \color[HTML]{9B9B9B}72.13 \\
     \midrule
     Chen et al.~\cite{chen2022self} & - & - & - & - & - & - & - & 71.79 & - & - & - & - & - & - & - & - & - \\
     FtEX~\cite{chen2023effective} & 76.70 & 69.30 & 72.49 & 70.69 & 87.49 & 70.96 & 76.89 & 74.52 & - & - & - & - & - & - & - & - & -\\
     Ours & 82.02 & 73.79 & 74.35 & 75.96 & 71.45 & 72.11 & 77.34 & \textbf{75.15} & 
     58.99 & 61.61 & 81.59 & 66.34 & 70.62 & 75.10 & 64.19 & \textbf{67.59} & \textbf{71.17} \\
    \bottomrule
  \end{tabular}
  }
\end{table*}

\begin{table*}[h]
  \caption{Performance comparison on the \textbf{DCASE 2022 dataset} between our proposed model and previous works. The pre-trained model is BEATs~\cite{beats}. Both fully-connected multi-branch LoRA and group adapter are applied for fine-tuning. All the results are reported using the harmonic mean of the AUC and pAUC across different machines following~\cite{dohi2022description}.}
  \label{tab:dcase22}
  \centering
  \adjustbox{width=1.0\linewidth,center=\linewidth}{
  \begin{tabular}{c|cccccccc|cccccccc|c}
    \toprule
    \multirow{2}*{Models} & \multicolumn{8}{c|}{\textbf{Development set}} & \multicolumn{8}{c|}{\textbf{Eval set}} & \textbf{All} \\
     & T.Car & T.Tra & Fan & G.box & Bearing & Slider & Valve & \textbf{Hmean} & T.Car & T.Tra & Fan & G.box & Bearing & Slider & Valve & \textbf{Hmean} & \textbf{Hmean} \\
     \midrule
     \color[HTML]{9B9B9B}2022 No.1~\cite{LiuCQUPT2022} & \color[HTML]{9B9B9B}76.07 & \color[HTML]{9B9B9B}71.77 & \color[HTML]{9B9B9B}68.63 & \color[HTML]{9B9B9B}84.38 & \color[HTML]{9B9B9B}77.90 & \color[HTML]{9B9B9B}86.37 & \color[HTML]{9B9B9B}91.37 & \color[HTML]{9B9B9B}78.77 & \color[HTML]{9B9B9B}86.13 & \color[HTML]{9B9B9B}67.05 & \color[HTML]{9B9B9B}57.34 & \color[HTML]{9B9B9B}77.29 & \color[HTML]{9B9B9B}63.27 & \color[HTML]{9B9B9B}73.86 & \color[HTML]{9B9B9B}80.77 & \color[HTML]{9B9B9B}70.97 & \color[HTML]{9B9B9B}74.67 \\ \midrule
     MFNV2~\cite{dohi2022description} & 53.91 & 51.54 & 58.19 & 59.37 & 58.70 & 53.18 & 62.28 & 56.74 & - & - & - & - & - & - & - & - & -\\
     AE~\cite{dohi2022description} & 57.68 & 50.16 & 60.21 & 62.14 & 54.19 & 59.30 & 50.55 & 56.32 & - & - & - & - & - & - & - & - & -\\
     STgram~\cite{liu2022anomalous} & 49.27 & 50.86 & 62.69 & 70.09 & 71.43 & 71.34 & 67.06 & 63.25 & - & - & - & - & - & - & - & - & -\\
     TWFRGMM~\cite{guan2023time} & 69.39 & 63.44 & 66.79 & 72.46 & 62.08 & 80.65 & 81.42 & 70.89 & - & - & - & - & - & - & - & - & - \\
     Ours & 77.34 & 77.96 & 63.3 & 70.89 & 58.59 & 79.71 & 87.91 & \textbf{72.43} & 50.93 & 50.68 & 52.64 & 86.4 & 64.08 & 65.09 & 79.59 & \textbf{61.69} & \textbf{66.63}  \\
    \bottomrule
  \end{tabular}
  }
\end{table*}

\begin{table*}[h]
  \caption{Performance comparison on the \textbf{DCASE 2023 dataset} between our proposed model and previous works. The pre-trained model is BEATs~\cite{beats}. Both fully-connected multi-branch LoRA and group adapter are applied for fine-tuning. All the results are reported using the harmonic mean of the AUC and pAUC across different machines following~\cite{dohi2023description}.}
  \label{tab:dcase23}
  \centering
  \adjustbox{width=1.0\linewidth,center=\linewidth}{
  \begin{tabular}{c|cccccccc|cccccccc|c}
    \toprule
    \multirow{2}*{Models} & \multicolumn{8}{c|}{\textbf{Development set}} & \multicolumn{8}{c|}{\textbf{Eval set}} & \textbf{All} \\
     & Bearing & Fan & G.box & Slider & T.Car & T.Tra & Valve & \textbf{Hmean} & B.Saw & Grinder & Shaker & T.Dro & T.Nsc & T.Tan & Vacuum & \textbf{Hmean} & \textbf{Hmean} \\
     \midrule
     \color[HTML]{9B9B9B}2023 No.1~\cite{JieIESEFPT2023} & \color[HTML]{9B9B9B}64.41 & \color[HTML]{9B9B9B}76.27 & \color[HTML]{9B9B9B}74.78 & \color[HTML]{9B9B9B}91.83 & \color[HTML]{9B9B9B}51.66 & \color[HTML]{9B9B9B}53.17 & \color[HTML]{9B9B9B}85.44 & \color[HTML]{9B9B9B}68.11 & \color[HTML]{9B9B9B}60.97 & \color[HTML]{9B9B9B}65.18 & \color[HTML]{9B9B9B}63.50 & \color[HTML]{9B9B9B}55.71 & \color[HTML]{9B9B9B}84.72 & \color[HTML]{9B9B9B}60.72 & \color[HTML]{9B9B9B}92.27 & \color[HTML]{9B9B9B}66.97 & \color[HTML]{9B9B9B}67.54 \\
     \midrule
     Han et al.~\cite{han2024exploring} & 57.10 & 62.76 & 67.52 & 79.11 & 63.47 & 57.35 & 67.79 & 64.31 & - & - & - & - & - & - & - & -  & - \\
     FeatEx~\cite{wilkinghoff2023self} & - & - & - & - & - & - & - & \textbf{66.95} & - & - & - & - & - & - & - & 68.52 & 67.73 \\
     AL~\cite{ho24_interspeech} & - & - & - & - & - & - & - & - & 63.03 & 66.03 & 63.00 & 59.59 & 75.19 & 67.74 & 97.44 & 68.49 & - \\
     Zhang et~al.~\cite{zhang2024dual} & - & - & - & - & - & - & - & - & - & - & - & - & - & - & - & 71.27 & - \\
     Ours & 71.01 & 57.70 & 62.65 & 85.46 & 57.49 & 62.85 & 60.36 & 64.25 & 68.15 & 63.52 & 85.16 & 62.96 & 85.60 & 70.75 & 89.93 & \textbf{73.70} & \textbf{68.65} \\
    \bottomrule
  \end{tabular}
  }
\end{table*}

\begin{table*}[h!]
  \caption{Performance comparison on the \textbf{DCASE 2024 dataset} between our proposed model and previous works. The pre-trained model is BEATs~\cite{beats}. Both fully-connected multi-branch LoRA and group adapter are applied for fine-tuning with DLCL loss. All the results are reported using the harmonic mean of the AUC and pAUC across different machines following~\cite{nishida2024description}. }
  \label{tab:dcase24}
  \centering
  \adjustbox{width=1.0\linewidth,center=\linewidth}{
  \begin{tabular}{c|cccccccc|cccccccccc|c}
    \toprule
    \multirow{2}*{Models} & \multicolumn{8}{c|}{\textbf{Development set}} & \multicolumn{10}{c|}{\textbf{Eval set}} & \textbf{All} \\
     & Bearing & Fan & G.box & Slider & T.Car & T.Tra. & Valve & \textbf{Hmean} & 3DPrin. & A.Com. & B.Mot. & H.Dry. & H.Dro. & R.Arm & Scan. & T.Bru. & T.Cir. & \textbf{Hmean} & \textbf{Hmean} \\
     \midrule
     \color[HTML]{9B9B9B}2024 No.1$^{*}$~\cite{LvAITHU2024} & \color[HTML]{9B9B9B}69.12 & \color[HTML]{9B9B9B}63.52 & \color[HTML]{9B9B9B}71.47 & \color[HTML]{9B9B9B}77.07 & \color[HTML]{9B9B9B}57.58 & \color[HTML]{9B9B9B}65.08 & \color[HTML]{9B9B9B}75.18 & \color[HTML]{9B9B9B}67.82 & \color[HTML]{9B9B9B}63.55 & \color[HTML]{9B9B9B}59.41 & \color[HTML]{9B9B9B}63.32 & \color[HTML]{9B9B9B}65.79 & \color[HTML]{9B9B9B}67.63 & \color[HTML]{9B9B9B}65.56 & \color[HTML]{9B9B9B}89.04 & \color[HTML]{9B9B9B}67.59 & \color[HTML]{9B9B9B}61.56 & \color[HTML]{9B9B9B}66.24 & \color[HTML]{9B9B9B}66.97 \\
     \color[HTML]{9B9B9B}2024 No.2$^{\dagger}$~\cite{JiangTHUEE2024}  & \color[HTML]{9B9B9B} 58.26  & \color[HTML]{9B9B9B} 66.32  & \color[HTML]{9B9B9B} 66.10  & \color[HTML]{9B9B9B} 61.63  & \color[HTML]{9B9B9B} 71.09  & \color[HTML]{9B9B9B} 79.85  & \color[HTML]{9B9B9B} 76.46  & \color[HTML]{9B9B9B} 67.78 &  \color[HTML]{9B9B9B} 60.53  & \color[HTML]{9B9B9B} 62.48  & \color[HTML]{9B9B9B} 62.40  & \color[HTML]{9B9B9B} 63.89  & \color[HTML]{9B9B9B} 67.21  & \color[HTML]{9B9B9B} 66.66  & \color[HTML]{9B9B9B} 86.97  & \color[HTML]{9B9B9B} 62.68  & \color[HTML]{9B9B9B} 61.96  & \color[HTML]{9B9B9B} 65.37   & \color[HTML]{9B9B9B} 66.40 \\
     \color[HTML]{9B9B9B}2024 No.3~\cite{ZhaoCUMT2024}  & \color[HTML]{9B9B9B} 52.50  & \color[HTML]{9B9B9B} 53.57  & \color[HTML]{9B9B9B} 55.09  & \color[HTML]{9B9B9B} 62.02  & \color[HTML]{9B9B9B} 69.73  & \color[HTML]{9B9B9B} 67.66  & \color[HTML]{9B9B9B} 51.84  & \color[HTML]{9B9B9B} 58.14  & \color[HTML]{9B9B9B} 53.68  & \color[HTML]{9B9B9B} 62.51  & \color[HTML]{9B9B9B} 71.74  & \color[HTML]{9B9B9B} 60.98  & \color[HTML]{9B9B9B} 65.50  & \color[HTML]{9B9B9B} 63.94  & \color[HTML]{9B9B9B} 72.70  & \color[HTML]{9B9B9B} 53.76  & \color[HTML]{9B9B9B} 58.79  & \color[HTML]{9B9B9B} 61.97  & \color[HTML]{9B9B9B} 60.23 \\
     \midrule
     AE~\cite{harada2023first} & 49.57 & 54.86 &	59.64 &	59.50 &	62.76 &	56.52 &	50.54 & 55.83 & 54.57 & 55.43 & 61.25 & 52.26 & 54.16 & 51.06 & 55.30 & 62.45 & 56.21 & 55.63 & 55.72 \\
     Ours & 64.19 & 62.78 & 65.36 & 60.77 & 59.01 & 61.40 & 78.80 & \textbf{64.11} & 60.94 & 60.04 & 68.37 & 67.68  & 66.22 & 65.59 & 85.13 & 68.15 & 65.81 & \textbf{66.95} & 65.50 \\
    \bottomrule
    \multicolumn{15}{l}{* The first place of DCASE 2024 is our team with ensemble pre-trained models and several strategies proposed in this paper.} \\
    \multicolumn{15}{l}{$\dagger$ The second place winner in the competition also followed our solution~\cite{han2024exploring,jiang24c_interspeech} and used ensemble pre-trained models.} \\
  \end{tabular}
  }
\end{table*}

\subsection{Evaluation of Dual-Level Contrastive Loss}
\label{sec:res_loss}
To evaluate the effectiveness of our proposed Dual-Level Contrastive Loss (DLCL) in the absence of attribute information, we conduct experiments using the DCASE 2024 dataset, and the results are presented in Table~\ref{tab:contras_loss}. In addition, we also implement the traditional offline clustering method for comparison. This traditional method consists of two stages: (1) the audio encoder is first fine-tuned with labeled data, and then it is used to extract representations for the unlabeled audio samples without the attributes. Finally, the static pseudo-labels are generated using a clustering algorithm; (2) these pseudo-labels are then employed to further fine-tune the encoder. This traditional method is denoted as PL in the experiments. 

The experimental results reveal that the traditional PL method with static pseudo-labels generated by the offline system are not sufficiently accurate, leading to an obvious performance decline compared to the baseline system. In contrast, our proposed DLCL achieves superior performance by dynamically selecting cluster centers via vector quantizers.
\begin{table}[h]
    \centering
    \caption{Performance comparison of fine-tuning BEATs~\cite{beats} with or without the proposed dual-level contrastive loss (DLCL) on the DCASE 2024 dataset. All hmean is the harmonic mean of both the Development and Evaluation sets. PL denotes the pseudo-labeling approach, which utilizes an offline clustering algorithm to generate static pseudo labels for fine-tuning.}
    \begin{tabular}{l|cc|c}
    \toprule
       \textbf{Models}  & \textbf{Dev} & \textbf{Eval} & \textbf{Hmean} \\ \midrule
       Baseline  & 62.42 & 64.46 & 63.42  \\
       \quad + PL & 60.51 & \textbf{65.55} & 62.51 \\
       \quad + DLCL & \textbf{64.88} & 65.08 & \textbf{64.98} \\
       \bottomrule
    \end{tabular}
    \label{tab:contras_loss}
\end{table}
To further validate the efficacy of the DLCL, the embeddings from several machines without attributes are selected, including ToyTrain, AirCompressor, BrushlessMotor, and HoveringDrone, and they are visualized using t-SNE, as shown in Fig.~\ref{fig:res_loss}(a) and (b). It shows the embeddings distributions from the models fine-tuned with normal AAM-Softmax classification loss and our proposed DLCL loss individually. It is observed that after fine-tuning with the proposed DLCL, audio samples from the same machine are dynamically clustered into distinct centers of the quantizer, which shows the advantage of the new loss. 
Furthermore, we don't use the attribute labels of HairDryer and RoboticArm to train the model to simulate the situation of missing labels, and visualize it using ground-truth labels in Fig.~\ref{fig:res_loss} (c) and (d). Compared to AAM-softmax, audio samples with DLCL from the same machine but with different attributes seem more discriminative, which is conducive to building a more robust machine sound encoder.

\subsection{Results Evaluation on DCASE 2020-2024 ASD Challenges}
\label{sec:res_overview}
In this section, we present a comprehensive evaluation of our proposed system on the five ASD challenge datasets from DCASE 2020 to 2024, as shown in Tables~\ref{tab:dcase20} to~\ref{tab:dcase24}.
The term “Ours” in the tables refers to our proposed anomalous sound detection framework, which is built upon the BEATs pre-trained model and integrates both the Fully-Connected Multi-Branch LoRA and the Machine-Aware Group Adapter in all experiments unless otherwise noted.
The DLCL algorithm is applied only in Table~\ref{tab:dcase24}, as it is specifically designed to address the absence of attribute labels during training, which occurs exclusively in the DCASE 2024 dataset.

For DCASE 2020~\cite{koizumi2020description}, the earliest and most widely used dataset, we selected all relevant published works for comparison, including the system that ranked first position in the competition. Regardless of advances in model architecture~\cite{Giri2020}, loss function~\cite{wilkinghoff2021sub, zeng2023joint}, data augmentation techniques~\cite{chen2023sw, zhang23fa_interspeech}, or other strategies~\cite{suefusa2020anomalous, dohi2021flow}, our proposed system using a pre-training model achieves the best performance on both the Development and Evaluation sets. Even when compared to systems that employ complex model ensembles, our system still has superior performance.

For DCASE 2021~\cite{kawaguchi2021description} and DCASE 2022~\cite{dohi2022description}, the primary challenge is from the domain shift, which increased the difficulty of anomalous sound detection. Shown as the results in Table \ref{tab:dcase21} and \ref{tab:dcase22}, it is observed that our system surpasses all previously published single system results, has a leading position in both the Development and Evaluation sets. Even when compared with the multi-system ensemble results of the competition champion, our method remains highly competitive.

The key change in DCASE 2023~\cite{dohi2023description} and DCASE 2024~\cite{nishida2024description} is that the machine types in the evaluation and training sets do not overlap, which was specifically designed to test the generalization ability on unseen machine sounds. As shown in Table~\ref{tab:dcase23}, while our system does not achieve the best results on the Development set, it significantly outperforms others on the Evaluation set, demonstrating the excellent generalization capability. 
For DCASE 2024, given that the dataset is relatively new and no relevant literature has been published, we compare our system with the top three ranked systems in the competition, and the results are listed in Table~\ref{tab:dcase24}. Among them, the first place is our team with several strategies proposed in this work, while the second place also followed our technical approach to adopt ensemble pre-trained models. 
It is observed that although our system only utilizes a single model, without employing complex strategies such as model ensemble or post-processing (e.g., SMOTE~\cite{chawla2002smote}), it can still achieve the comparable performance to the top-performing system, demonstrating the strong ability of our proposed approaches. And compared to other solutions that do not leverage pre-trained models, such as the method ranked third in the challenge, our approach demonstrates significantly improved performance.

\section{conclusion}
In this paper, we have introduced a robust machine anomalous sound detection (ASD) model that leverages self-supervised pre-trained models to enhance generalization performance. Despite the inherent inconsistencies between the pre-training datasets and the ASD task, our findings demonstrate that pre-training provides substantial benefits, facilitating knowledge transfer from large-scale speech and audio corpora. To address overfitting and knowledge retention when fine-tuning with limited data, we explored Fully-Connected Low-Rank Adaptation (LoRA) as an efficient alternative to full fine-tuning. Furthermore, we proposed the Machine-aware Group Adapter module, which effectively captures inter-machine variations within a unified framework, leading to improved generalization across diverse machine types. Additionally, to overcome the challenge of missing attribute labels, we designed a novel objective function that dynamically clusters unattributed data using vector quantization and optimizes learning via a dual-level contrastive loss. Extensive experiments conducted on benchmark datasets from DCASE 2020 to 2024 validate the effectiveness of our approach. The results demonstrate consistent and significant improvements over existing methods, highlighting the efficacy of the proposed methods, i.e. self-supervised pre-training, LoRA-based adaptation, machine-aware modeling, and contrastive clustering. These findings suggest promising directions for future research, including the exploration of more advanced self-supervised learning techniques and domain adaptation strategies to further enhance the robustness of ASD systems in real-world applications.


\ifCLASSOPTIONcaptionsoff
  \newpage
\fi

\bibliographystyle{IEEEtran}

\bibliography{mybib}

\end{document}